\documentclass[aps,showkeys,nofootinbib,nopacs,12pt]{revtex4-1}

\usepackage{amsmath}
\usepackage[titletoc,title]{appendix}
\usepackage{graphicx,epic,eepic,epsfig,amsmath,latexsym,amssymb,verbatim,color}
\usepackage{dsfont}
\usepackage{MnSymbol}
\usepackage{tikz}
\usepackage{amssymb}
\usepackage{graphicx}
\usepackage{url}
\usepackage{hyperref}
\hypersetup{colorlinks=true,citecolor=blue,linkcolor=blue,filecolor=blue,urlcolor=blue,breaklinks=true}

\usepackage{epstopdf}
\usepackage{amsmath}
\usepackage{stmaryrd}
\usepackage{amsthm}

\usepackage{algorithm}
\usepackage[noend]{algpseudocode}

\newcommand{\cE}{\mathcal{E}}

\newcommand{\cH}{\mathcal{H}}

\newcommand{\llrr}[1]{\llbracket #1 \rrbracket} %[[#1]]

\newcommand{\bra}[1]{\ensuremath{\left\langle#1\right|}}
\newcommand{\ket}[1]{\ensuremath{\left|#1\right\rangle}}
\newcommand{\op}[2]{\ensuremath{|#1 \rangle \langle #2 |}}

\newcommand{\compilername}{$Q|SI\rangle $\,}
\newcommand{\cwhile}{\mathbf{while}}
\newcommand{\cif}{\mathbf{if}}
\newcommand{\compactT}{\langle T \rangle}
\newcommand{\liquid}{LIQU$i|\rangle$} %liquid name

\newtheorem{definition}{Definition}[section]

\begin{document}
    \title{\compilername: A Quantum Programming Environment}

\author{Shusen Liu}
\email{Shusen.Liu@student.uts.edu.au}
\affiliation{%
Centre for Quantum Software and Information, Faculty of Engineering and Information Technology, University of Technology Sydney, NSW 2007, Australia
}%

\author{Xin Wang}%
\affiliation{%
    Centre for Quantum Software and Information, Faculty of Engineering and Information Technology, University of Technology Sydney, NSW 2007, Australia
}%

\author{Li Zhou}%
\affiliation{%
Centre for Quantum Software and Information, Faculty of Engineering and Information Technology, University of Technology Sydney, NSW 2007, Australia
}%
\author{Ji Guan}%
\affiliation{%
Centre for Quantum Software and Information, Faculty of Engineering and Information Technology, University of Technology Sydney, NSW 2007, Australia
}%

\author{Yinan~Li}%
\affiliation{%
    Centre for Quantum Software and Information, Faculty of Engineering and Information Technology, University of Technology Sydney, NSW 2007, Australia
}%
%\\
\author{Yang He}%
\affiliation{%
Centre for Quantum Software and Information, Faculty of Engineering and Information Technology, University of Technology Sydney, NSW 2007, Australia
}%

\author{Runyao Duan}
\email{Runyao.Duan@uts.edu.au}
\affiliation{%
Centre for Quantum Software and Information, Faculty of Engineering and Information Technology, University of Technology Sydney, NSW 2007, Australia
}%

\author{Mingsheng Ying}
\email{Mingsheng.Ying@uts.edu.au}
\altaffiliation[Also at ]{Department of Computer Science and Technology, Tsinghua University, Beijing, China}
\altaffiliation[Also at ]{State Key Laboratory of Computer Science, Institute of Software, Chinese Academy of Sciences, Beijing, China}
\affiliation{%
Centre for Quantum Software and Information, Faculty of Engineering and Information Technology, University of Technology Sydney, NSW 2007, Australia
}%

\begin{abstract}
This paper describes a quantum programming environment, named \compilername\footnote{The software of \compilername is available at \url{http://www.qcompiler.com}.}. It is a platform embedded in the .Net language that supports quantum programming using a quantum extension of the $\cwhile$-language. The framework of the platform includes a compiler of the quantum $\cwhile$-language and a suite of tools for simulating quantum computation, optimizing quantum circuits, and analyzing and verifying quantum programs. Throughout the paper, using \compilername to simulate quantum behaviors on classical platforms with a combination of components is demonstrated. The scalable framework allows the user to program customized functions on the platform. The compiler works as the core of \compilername bridging the gap from quantum hardware to quantum software. The built-in decomposition algorithms enable the universal quantum computation on the present quantum hardware.  
\end{abstract}

 
\maketitle

\section{Introduction}
It is well-known that quantum computers can solve certain categories of problems much more efficiently than classical computers; for example, Shor's factoring algorithm~\cite{Shor1997},  Grover's search algorithm~\cite{grover1996fast} and more recently Harrow, Hassidim and Lloyd's algorithm for systems of linear equations~\cite{harrow2009}. In recent years, governments and industries around the globe have been racing to build quantum computers. As happened in the history of classical computing, once quantum computers are commercialized, programmers will certainly need a modern platform that can express and implement quantum algorithms without considering the trivialities of their circuits. 
Such a platform will be even more helpful for quantum programming than in classical computing because physically implement quantum algorithms in a quantum system is somewhat counterintuitive. Using a platform like \compilername could help programmers understand some of these features, which may help to (partially) avoid some of the errors.  
 
Several quantum programming platforms have been developed in the last two decades.  
The first quantum programming language, QCL, was proposed by \"{Omer}~\cite{qcl1,qcl2} in 1998. It was implemented in C++. 
A very similar quantum programming language, Q language, was defined by Bettelli et al.~\cite{qlanguage} in 2003, which was implemented as a C++ library. 
In 2000, qGCL was introduced by Sanders and Zuliani~\cite{Sanders2000} as a quantum extension of GCL (Dijkstra's Guarded-Command Language) and pGCL (a probabilistic extension of GCL). Over the last few years, some more scalable and robust quantum programming platforms have emerged: A scalable functional programming language Quipper for quantum computing was proposed by Green et al.~\cite{Green2013} in 2013. This was implemented using Haskell as its host language. 
\liquid was developed in 2014 by Wecker and Svore from QuArc (the Microsoft Research Quantum Architecture and Computation team)~\cite{Liquid} as a modern tool-set and is embedded in another functional programming language F\#. In the same year, the quantum programming language Scafford was defined by JavadiAbhari et al.~\cite{Scafford}. Its compilation system ScaffCC was developed in the article~\cite{scaffCC}.  
Smelyanskiy et al.~\cite{intel} at Intel built a parallel quantum computing simulator qHiPSTER that can simulate up to $40$ qubits on a supercomputer with very high performance.

{\vskip 4pt}
    
\textbf{Contributions of this paper}: This paper presents a powerful and flexible new quantum programming environment called \compilername\footnote{\url{http://www.qcompiler.com}}, named after our research center\footnote{\url{http://www.qsi.uts.edu.au}}. The \textit{core} of \compilername is \textit{a quantum programming language} and \textit{its compiler}. This language is a quantum extension of the $\cwhile$-language. It was first defined in~\cite{ying2011floyd} along with a careful study of its operational and denotational semantics (see also~\cite{Ying2016}, Chapter 3). The language includes a measurement-based case statement and a measurement-based $\cwhile$-loop. These two program constructs are extremely convenient for describing large-scale quantum algorithms, e.g., quantum random walk-based algorithms. 

For operations with quantum hardware, we have defined a new assembly language called f-QASM (Quantum Assembly Language with feedback) as an interactive command set. f-QASM is an extension of the instruction set QASM (Quantum Assembly Language) introduced in~\cite{svore2006layered}. A feedback instruction has been added that allows the efficient implementation of measurement-based case and loop statements. A compiler then transforms the quantum $\cwhile$-program into a sequence of f-QASM instructions and further generates a corresponding quantum circuit equivalent to the program (i.e., a sequence of executable quantum gates). \compilername also has a module for optimizing the quantum circuits as well as a module to simulate its quantum programs on a classical computer. 
Two novel features set \compilername apart from other existing quantum programming environments:
\begin{itemize} 
\item \textit{A quantum program analyzer}. Several algorithms for termination analysis and for computing the average running time of quantum programs were developed by one of the authors in~\cite{ying2010quantum,ying2013verification}. In addition, a semi-definite programming (SDP) algorithm generates invariants of quantum $\cwhile$-loops was developed by one of the authors in~\cite{ying2017invariants}. These algorithms have been implemented in \compilername for the static analysis of quantum programs. In turn, this program analyzer helps the compiler to optimize the implementation of quantum programs.  
\item \textit{A quantum program verifier}. A logic in the Floyd-Hoare style was established in~\cite{ying2011floyd} (see also~\cite{Ying2016}, Chapter 4). This logic, which reasons about the correctness of quantum programs, has been written in the quantum $\cwhile$-language. Recently, a theorem prover was implemented by Liu et al.~\cite{liu2016theorem} for quantum Floyd-Hoare logic based on Isabelle/HOL.
We plan to link \compilername with the quantum theorem prover presented in~\cite{liu2016theorem} and provide this facility in our platform for the verification of quantum programs. 
\end{itemize}

\section{Quantum $\cwhile$-Language}
For convenience, a brief review of the quantum $\cwhile$-language follows. The quantum $\cwhile$-language is a pure quantum language without classical variables. It assumes only a set of quantum variables denoted by the symbols $q_0, q_1, q_2,...$.  However, in practice, almost all existing quantum algorithms involve elements of both classical and quantum computation. Therefore, \compilername has been designed such that the quantum $\cwhile$-language can be embedded into C\#, which brings a significant level of convenience to program design. Some explanations of the quantum program constructs follow; For more detailed descriptions and examples, see~\cite{ying2011floyd} and Chapter 3 of~\cite{Ying2016}.   
The quantum $\cwhile$-language is generated using the following simple syntax:
\[
\begin{split}
\mathbf{S}::=\mathbf{skip}\ &|\ q:=\ket{0}\ |\ \bar{q}=U[\bar{q}]\ |\ S_1;S_2\ |\mathbf{if}\ (\square m\cdot M[\bar{q}]=m\to S_m)\ \mathbf{fi}\\
&|\ \mathbf{while}\ M[\bar{q}]=1\ \mathbf{do}\,\mathbf{S}\, \mathbf{od}.
\end{split}
\]

\begin{description}
    \item[Skip] As in the classical $\cwhile$-language, the statement $\mathbf{skip}$ does nothing and terminates immediately.
    \item[Initialization] The initialization statement ``$q:=\ket{0}$" sets the quantum variable $q$ to the basis state $\ket{0}$.
    \item[Unitary transformation]
    The statement  ``$\bar{q}:=U[\bar{q}]$" means that a unitary transformation (quantum gate) $U$ is performed on quantum register $\bar{q}$ leaving the other variables unchanged. 
    \item[Sequential composition] 
     As in a classical programming language, in the composition $S_1;S_2$, program $S_1$ is executed first. Once $S_1$ terminates, $S_2$ is executed. 
     \item[Case statement]  
    In the case statement $\cif\ (\square m\cdot M[\bar{q}]=m\to S_m)\ \mathbf{fi}$, $M$ is a quantum measurement with $m$ representing its possible outcomes. To execute this statement, $M$ is first performed on the quantum register $\bar{q}$ and a measurement outcome $m$ is obtained with a certain probability. Then, the subprogram $S_m$ is selected according to the outcome $m$ and executed. The difference between a classical case statement and a quantum case statement is that the state of the quantum program variable $\bar{q}$ is changed after performing the measurement. 
    \item[$\cwhile$-Loop]  In the loop $\mathbf{while}\ M[\bar{q}]=1\ \mathbf{do}\,\mathbf{S}\, \mathbf{od}$, $M$ is a ``yes-no" measurement with only two possible outcomes: $0$ and $1$. During execution, $M$ is performed on the quantum register $\bar{q}$ to check the loop guard. If the outcome is $0$, the program terminates. If the outcome is $1$ the program executes the loop body $S$ and continues. Note that here the state of the program variable $\bar{q}$ is also changed after measuring $M$.  
\end{description}

\section{The Structure of \compilername}
This section provides an introduction to the basic structure of \compilername, leaving the details to be described in subsequent sections. \compilername is designed to offer a unified general-purpose programming environment to support the quantum $\cwhile$-language. It includes a compiler for quantum $\cwhile$-programs, a quantum computation simulator, and a module for the analysis and verification of quantum programs. We have implemented \compilername as a deeply embedded domain-specific platform for quantum programming using the host language C\#. 

\begin{figure}  
    \centering  
    \includegraphics[width=165mm]{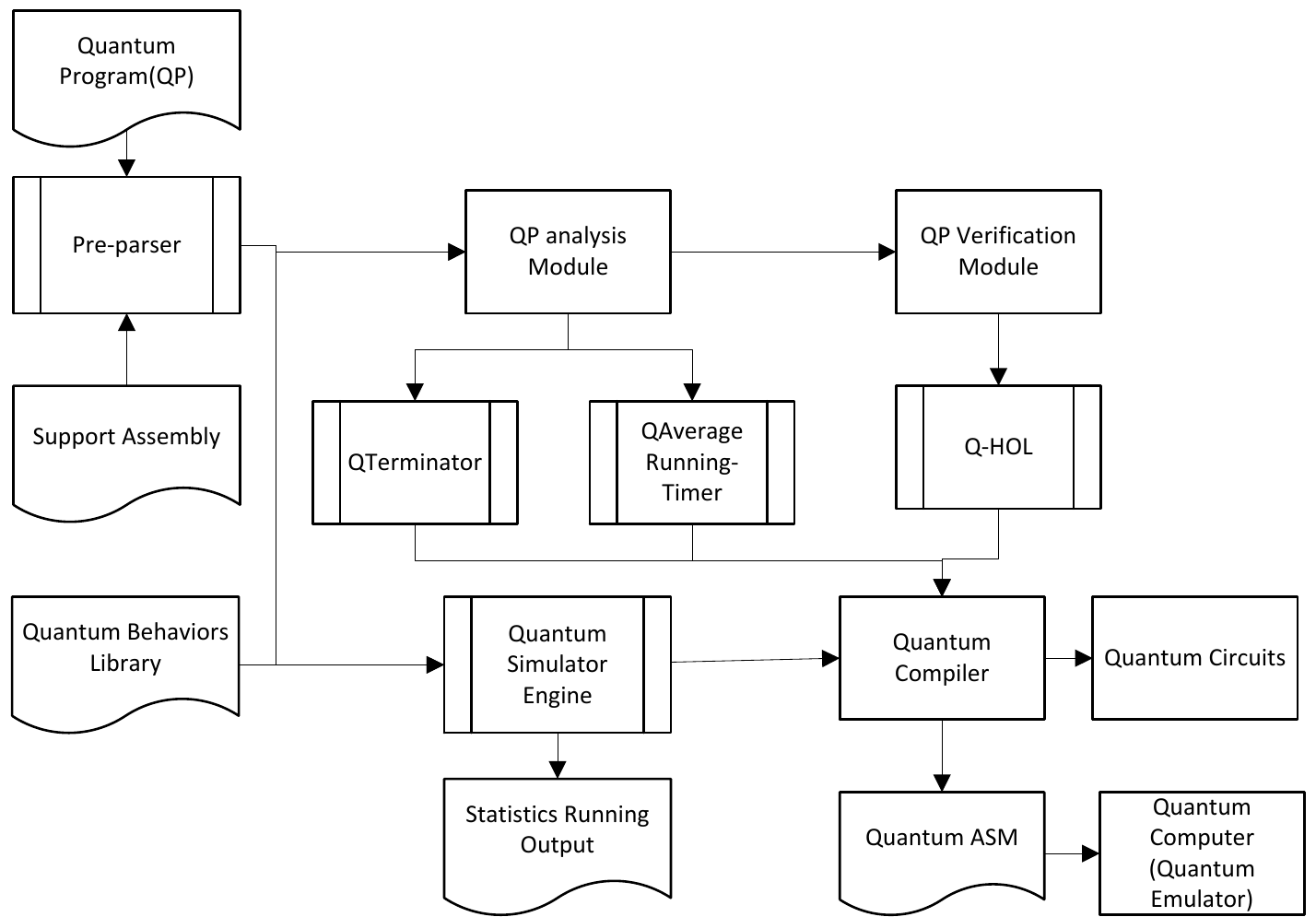}  
    \caption{Framework of \compilername }  
    \label{fig:framework}  
\end{figure}

\compilername's framework is shown in the Fig.~\ref{fig:framework}.

\subsection{Basic features of \compilername}
The main features of \compilername are explained as follows:
\begin{description}
    \item[Language supporting] \compilername is the first platform to support the quantum $\cwhile$-language. In particular, it allows users to program with measurement-based case statements and $\cwhile$-loops. The two program constructs provide more efficient and clearer descriptions of some quantum algorithms,  such as quantum walks and Grover's search algorithm.
    \item[Quantum type enriched] Compared to other simulators and analysis tools, \compilername supports quantum types beyond pure qubit states, such as density operators, mixed states, etc. These types have unified operations and can be used in different scenarios. This feature provides high flexible usability and facilitates the programming process. 
    \item[Dual mode] \compilername has two executable modes. ``Running-time execution" mode simulates quantum behaviors in one-shot experiments. ``Static execution" mode is mainly designed for quantum compilation, analysis, and verification.
    \item[f-QASM instruction set] Defined as an extension of Quantum Assembly Language (QASM)~\cite{svore2006layered}, f-QASM is essentially a quantum circuit description language that can be adapted for a variety purpose. In this language, every line has only one command. f-QASM's `goto' structure contains more information than the original QASM~\cite{svore2006layered} or space efficient QASM-HL~\cite{scaffCC}. f-QASM can also be used for further optimization and analysis.    
    \item[Quantum circuits generation] Similar to modern digital circuits in classical computing, quantum circuits provide a low-level representation of quantum algorithms~\cite{svore2006layered}. Our compiler can produce a quantum circuit from a program written in the high-level quantum $\cwhile$-language.  
        \item[Arbitrary unitary operator implementation] The \compilername platform includes the Solovay-Kitaev algorithm~\cite{dawson2005solovay} together with two-level matrix decomposition algorithm~\cite{Nielsen2000} and a quantum multiplexor (QMUX) algorithm~\cite{ShendeBullockMarkov2006}. Therefore, an arbitrary unitary operator could be transferred into a quantum circuit consisting of quantum gates from a small pre-defined set of basic gates once these are available from quantum chip manufactures.  
    \item[Gate-by-gate description] Similar to other quantum simulators, \compilername has a gate-by-gate description feature. Some basic quantum gates are provided inherently in our platform. Users can use them to build their desired quantum circuits gate-by-gate. We have also provided a decomposition function to generate arbitrary two-dimensional controlled-unitary gates for emulation feasibility.
\end{description}

\subsection{Main components of \compilername}
The \compilername platform mainly consists of four parts.
\begin{description}
    \item[Quantum Simulation Engine] This component includes some support assemblies, a quantum mechanics library and a quantum simulator engine. The support assemblies provide supporting for the quantum types and quantum language. More specifically, they provide a series of quantum objects, and reentrant encapsulated functions to play the role of the quantum program constructs $\cif$ and $\cwhile$. The quantum mechanics library provides the behaviors for quantum objects such as unitary transformation and measurement including the result and post-state. The quantum simulator engine is designed as an execution engine. It accepts quantum objects and their rules from the quantum mechanics library and converts them into probability programming which can be executed on a classical computer.
 
\item[Quantum Program (QP) Analysis Module] This module currently contributes two sub-modules to static analysis mode: the ``QTerminator" and the ``QAverage Running-Timer". The former provides the terminating information, and the latter evaluates the running time of the given program. Their outputs are sent to the quantum compiler at the next stage for further usage. 

\item[QP Verification Module] This module is a tool for verifying the correctness of quantum programs. It is based on quantum Hoare logic, which was introduced by one of the authors in~\cite{ying2011floyd} and is still under development. One possibility for its future advancement is to link \compilername to the quantum theorem prover developed by Liu et al~\cite{liu2016theorem}. 

\item[Quantum Compiler] The compiler consists of a series of tools to map a high-level source program representing a quantum algorithm into a quantum device related language~\cite{svore2006layered}, e.g., f-QASM and further into a quantum circuit. Our target is to be able to implement any source code without considering the details of the devices. It will ultimately run on, i.e., to automatically construct a quantum circuit based on the source code. A tool to optimize the quantum circuits will be added to the compiler in the future.
\end{description}

\subsection{Implementation of \compilername}
\begin{figure}[!t]
%    \centering  
    \includegraphics[width=172mm]{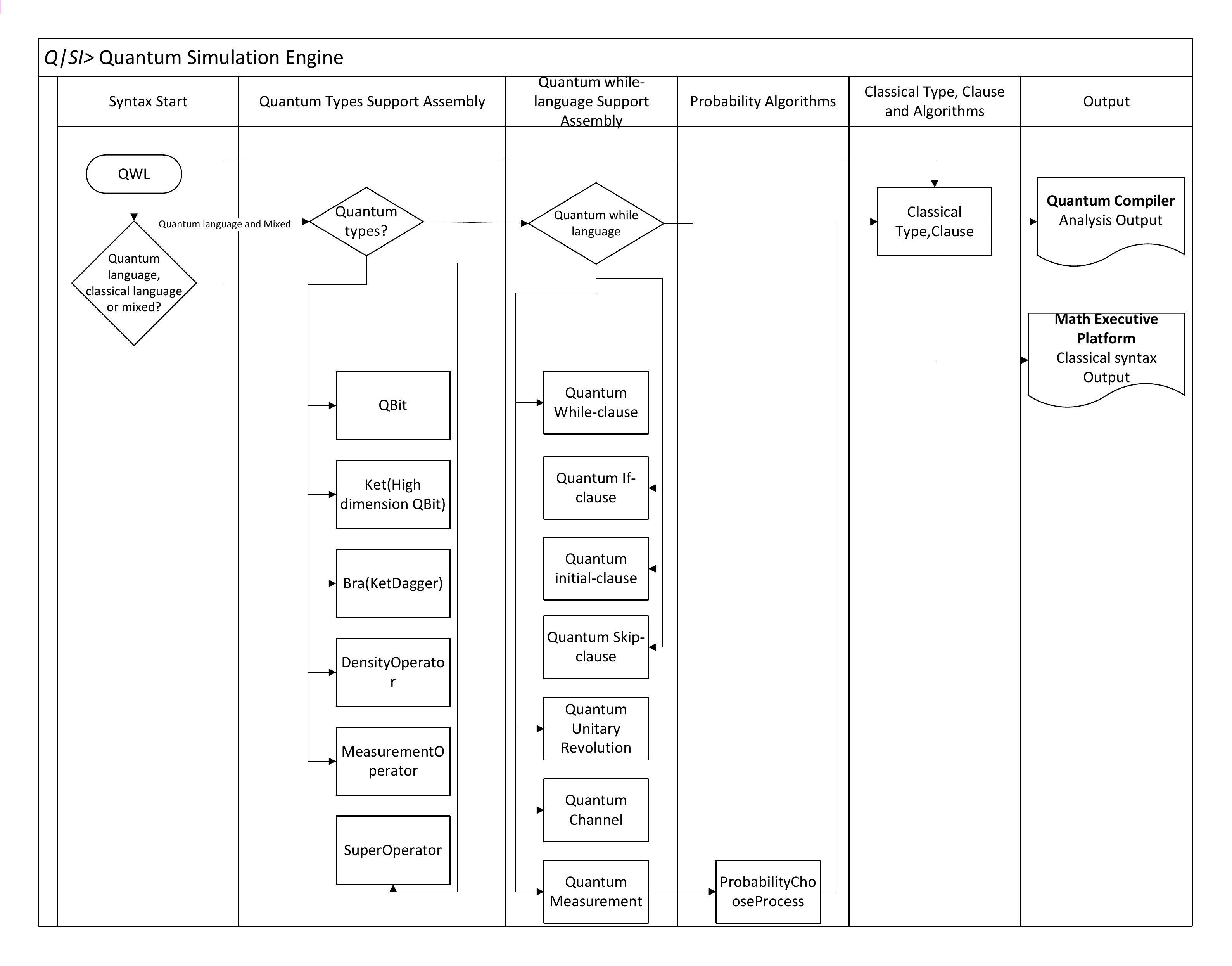}
    \caption{\compilername\,Pre-parser}  
    \label{fig:QWL}  
\end{figure}

One of the basic problems during implementation is how to use probabilistic and classical algorithms to simulate quantum behaviors. To support quantum operations, \compilername has been enriched with data structures from a quantum simulation engine. Fig.~\ref{fig:QWL} shows the procedure for simulating a quantum engine. 
Three types of languages are supported: pure quantum $\cwhile$-language, classical $\cwhile$-language and a mixed language. The engine starts a support flow path when it detects the quantum part of a program. Then, the engine checks the quantum type for each variable and operator and executes the corresponding support assembly. As mentioned, one of the main features of \compilername  is that it supports programming in the quantum $\cwhile$-language. This feature is provided by the quantum $\cwhile$-language support assemblies. All of the quantum behaviors are explained by probabilistic algorithms on a classical computer. The outputs are extended C\# languages which can be run on a .Net framework or can be explained in f-QASM and quantum circuits by the compiler.  

The quantum simulation engine involves numerous matrix computations and operations. In \compilername, Math.net is used for matrix computation. Math.NET is an open-source initiative to build and maintain toolkits that cover fundamental mathematics. It targets both the everyday and the advanced needs of .Net developers\footnote{\url{https://www.mathdotnet.com}}. It includes numerical computing, computer algebra, signal processing and geometry. Math.net is also able to accelerate matrix calculations when the simulation includes a MIC (Many Integrated Core Architecture) device.

In static analysis mode, Roslyn is our chosen auxiliary code analysis tool. Roslyn is a set of open-source compilers and code analysis APIs for C\# and Basic languages. Since our platform is embedded in the .Net framework for the C\# language, Roslyn is used as a parser to produce an abstract syntax tree (AST) for further analysis.

\section{The Quantum Compiler}
A compiler works as a connection between different devices and data structures and serves several different functions. It produces f-QASM code, which can be used to emulate a real or virtual quantum processor. It provides quantum circuits for quantum chip design. It also optimizes quantum circuits. 

The \compilername compiler is heavily dependent on other modules. It collects data structures from the quantum simulation engine and splits the program into several parts: variables, quantum gates, quantum measurements, entry and exit points for each clause with their positions. It constructs an AST (Abstract Syntax Tree) from the program, then reconstructs the program as a sequence of f-QASM instructions for further use. Based on f-QASM, the compiler provides a method for decomposing the unitary operators. It can decompose an arbitrary unitary operator $U(n)$ into a sequence of basic quantum gates from a pre-defined set $\{U_1,U_2,\ldots,U_m\}$ where $U_1,U_2,\ldots,U_m \in U(2)$ (qubit gate). This corresponds to a scenario in quantum device development: people need universal computation in spite of only a few of gates, which can be produced by the manufacturers. Further, the quantum $\cwhile$-language delivers the power of loops, but it also increases the complexity of compilation. A quantum program with a loop structure is much harder to trace than the one without loops. The QP Analysis module provides static analysis tools including a ``QTerminator'' for termination checking and a ``QAverage Running-Timer'' for computing the expected running time. In addition, the QP Verification module, still in development, is being designed to verify quantum programs. Once complete, programmers will be able to insert to debug program behaviors.

\subsection{f-QASM}
QASM (Quantum Assembly Language) is widely used in modern quantum simulators. It was first introduced in~\cite{svore2006layered} and is defined as a technology-independent reduced-instruction-set computing assembly language extended by a set of quantum instructions based on the quantum circuit model. The article~\cite{ying2011flowchart} carefully characterizes its theoretical properties. In 2014, A.JavadiAbhari et al.~\cite{scaffCC} defined a space-consuming flat description and denser hierarchical description QASM, called QASM-HL. Recently, Smith et al.~\cite{Quil} proposed a hybrid QASM for classical-quantum algorithms and applied it in Quil. Quil is the front-end of Forest which is a tool for quantum programming and simulation that works in the cloud.

We propose a specific QASM format, called f-QASM (Quantum Assembly Language with feedback). The most significant motivation behind our variation is to translate the inherent logic of quantum program written in a high-level programming language into a simple command set, i.e., so there is only one command in every line or period. However, a further motivation is to solve an issue raised by the IBM QASM 2.0 list and provide the ability to have conditional operations to implement feedback based on measurement outcomes.
\subsection{Basic definition of f-QASM}
Let us first define the registers: 
\begin{itemize}
    \item Define $\{r_1,r_2,\ldots\}$ as the finite classical registers.
    
    \item Define $\{q_1,q_2,\ldots\}$ as the finite quantum registers.
    
    \item Define $\{fr_1,fr_2,\ldots\}$ as the finite flag registers. These are a special kind of classical registers that are often used to illustrate partial results of the code segment. In most cases, the flag registers can not be operated directly by any users code.
\end{itemize}

Then we define two kinds of basic operations: 
\begin{itemize}
    \item Define the command  ``$op(q)$'' as $q:=op(q)$, where $op$ is a unitary operator and $q$ is a quantum register. 
    
    \item Define the command ``$\{op\}(q)$" as $r:=\{op\}(q)$, where $\{op\}$ is a set of measurement operators, $q$ is a quantum register, and $r$ a is classical register.\end{itemize}

After defining registers and operations, we can define some assembly functions:
\begin{itemize}
    \item Define ``$INIT(q)$" as $q:=\ket{0}\bra{0}$, where $q$ is a quantum register. The value of $q$ is assigned into 
    $\begin{pmatrix}
    1&0\\
    0&0
    \, 
    \end{pmatrix}$.
    
    \item Define ``$OP\{q,num\}$", where $q$ is a quantum register, $num \in \mathbb{N}$ and $OP$ is an operator, in another functional form of $q:=op(q)$.  When $num$ is $0$, it means the unitary operator belongs to the pre-defined set of basic quantum gates which can be prepared by the manufacturer or the user. Otherwise, $num$ can only be used after being decomposed into basic gates, or be ignored.
    
    \item Define ``$MOV(r_1,r_2)$", $r_1$ and $r_2$ are the classical registers. This function assigns the value of the register $r_2$ to the register $r_1$ and empties $r_2$. 
    
    \item Define ``$CMP(r_1,r_2)$" as $fr_1=\delta(r_1,r_2)$ or as $fr_1=(r_1==r_2)$, where $r_1,r_2$ are two classical registers, $\delta$ is the function comparing whether $r_1$ is equal to $r_2$: if $r_1$ is equal to $r_2$ then $fr_1=1$; otherwise $fr_1=0$.
    
    \item Define $JMP\ l_0$ as the current command goes to the line indexed by $l_0$.
    
    \item Define $JE \ l_0$ as indexing the value of $fr_1$ and jumping. If $fr_1$ is equal to $1$ then the compiler executes $JMP \ l_0$, otherwise it does nothing.
\end{itemize}

\subsection{f-QASM examples}
Some simple examples to help readers understand f-QASM follow, 
\subsubsection{Initialization}
$\mathbf{q:=\ket{0}}$ means the program initializes the quantum register $q$ in the state $\ket{0}$. In f-QASM, initializing two quantum registers $Q1$ and $Q2$ in the state $\ket{0}$ would be written as
\begin{verbatim}
INIT(Q1);
INIT(Q2);
\end{verbatim}

\subsubsection{Unitary transformation}
$\mathbf{\bar{q}=U[\bar{q}]}$ means the program performs a unitary transformation on the register $q$. The compiler will check whether the unitary matrix is a basic gate. An example program segment of unitary transformation follows: 
\begin{verbatim}
hGate(q1);
\end{verbatim}
Here we support $\emph{hGate}$ is a Hadamard gate performed on single qubit, i.e., $hGate=\frac{1}{\sqrt{2}}\begin{pmatrix}
1&1 \\
1&-1
\,
\end{pmatrix}\,.$ To transform this into an f-QASM instruction, it is written as 
\begin{verbatim}
hGate(q1, 0);
\end{verbatim}

\subsubsection{Case statement}
The following program segment is written as a case statement in quantum $\cwhile$-language: 
\begin{verbatim}
QIf(m(q1)
() =>
{
xGate(q1);
},
() =>
{
hGate(q1);
}
);
zGate(q1);
\end{verbatim}
where $\emph{hGate}$ is a Hadamard gate performed on single qubit, $\emph{xGate}$ is a bit-flip gate performing on single qubit $xGate=\begin{pmatrix}
0&1 \\
1&0
\,
\end{pmatrix}$, and $\emph{zGate}$ is a phase-flip gate $zGate=\begin{pmatrix}
1&0 \\
0&-1
\,
\end{pmatrix}$. Here we assume that all the gates can be provided. $M$ is a user-defined measurement.
The compiler interprets this segment as the following f-QASM instructions: 
\begin{verbatim}
MOV(r,{M}(q1));
CMP(r,0);
JE L1;
CMP(r,1);
JE L2;
L1:
xGate(q1,0);
JMP L3;
L2:
hGate(q1,0)
JMP L3;
L3:
zGate(q1,0);
\end{verbatim}

\subsubsection{Loop}
A loop construct is provided using $QWhile(M(q))$, where $QWhile$ is a key word, $M$ is a measurement and $q$ is a quantum register.
An example program segment with quantum $\cwhile$-loop follows:
\begin{verbatim}
QWhile(m(q1),
() =>
{
xGate(q1);
}
);
hGate(q1);
\end{verbatim} 

Both $\emph{hGate}$ and $\emph{xGate}$ are basic gates which have been described above. It could be transformed into f-QASM as follows: 

\begin{verbatim}
L1:
MOV(r,{M}(q1));
CMP(r,0);
JE L2;
XGate(q1,0);
JMP L1;
L2:
hGate(q1,0);
\end{verbatim}

\subsection{Decomposition of a general unitary transformation}
Given a set $\{U_1,U_2,\ldots,U_n\}$ of basic gates. If any unitary operator can be approximated to arbitrary accuracy by a sequence of gates from this set, then the set is said to be universal~\cite{Nielsen2000}.

In the compiler, there are two kinds of built-in decomposition algorithms. One is the QR method given in~\cite{barenco1995elementary,Nielsen2000}. It consists of the following steps:
\begin{enumerate}
    \item An arbitrary unitary operator is decomposed exactly into (the composition of) a sequence of unitary operators that act non-trivially only on a subspace spanned by two computational basis states; 
    \item Each unitary operator, which only acts non-trivially on a subspace spanned by two computational basis states are further expressed using single qubit gates ($U(2)$) and the CNOT gate;
    \item Each single qubit gate can be decomposed into a sequence of gates from a given small set of basic (single qubit) gates using the Solovay-Kitaev theorem~\cite{dawson2005solovay}.
\end{enumerate}
The other is the QSD method presented in~\cite{ShendeBullockMarkov2006}, consisting of the following steps: 
\begin{enumerate}
    \item An arbitrary operator is decomposed into three multiplexed rotations and four generic $U(2^{d-1})$ operators, where $d$ is the number of qubits;
    \item Repeatedly execute step 1 until $U(4)$ is generated;
    \item The $U(4)$ operator is decomposed into $U(2)$ operators with two extra CNOT gates;
    \item Each single qubit gate in $U(2)$ is decomposed into gates from a given small set of basic (single qubit) gates using the Solovay-Kitaev theorem~\cite{dawson2005solovay}.
\end{enumerate}
 
\section{The Quantum Simulator}
\subsection{Quantum types} 
\begin{figure}[!pb]  
    \centering  
    \includegraphics[width=165mm]{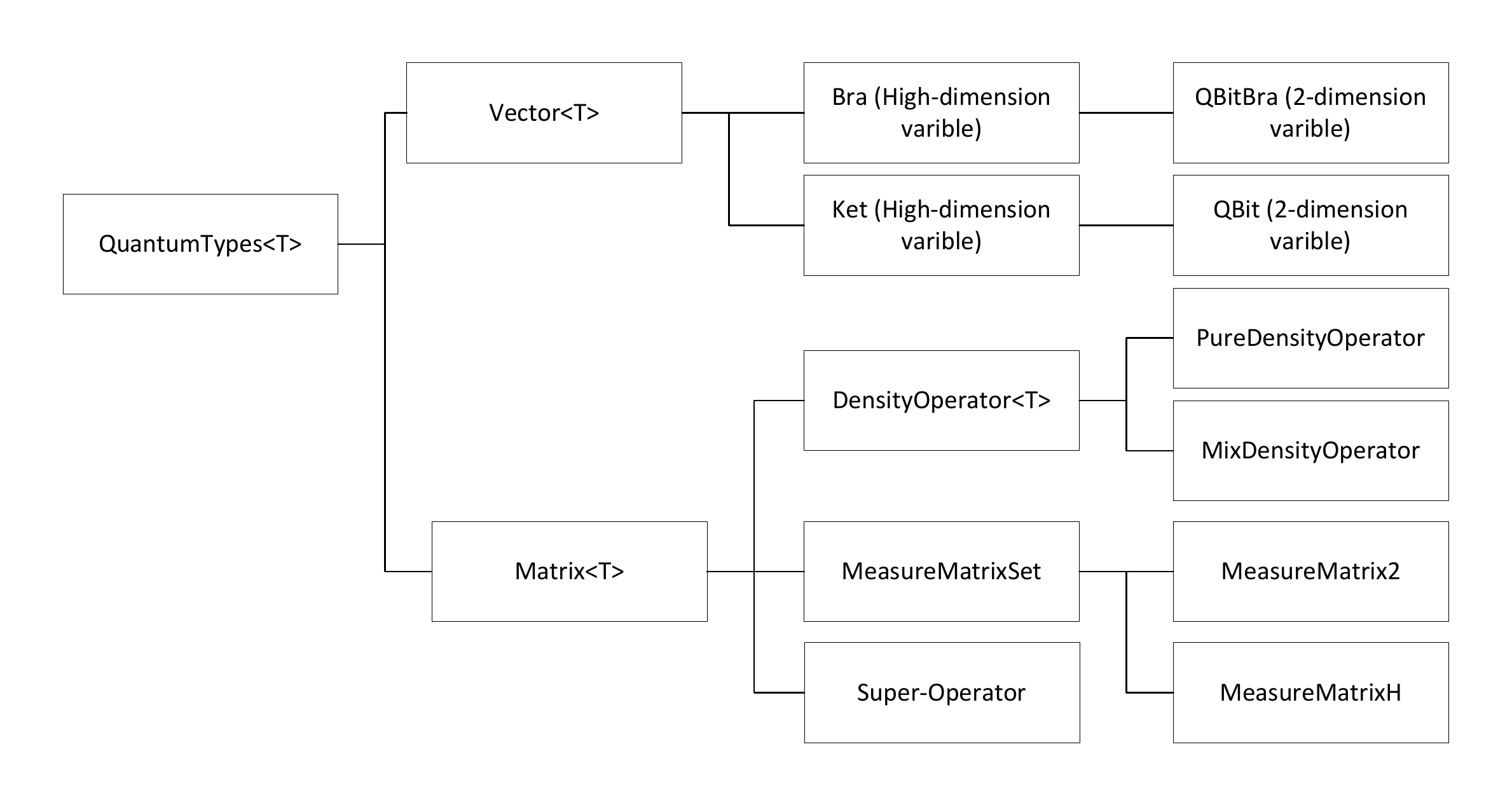}  
    \caption{\compilername\,Quantum types layer}  
    \label{fig:quantum_types_layer}  
\end{figure}

Data types can be extended from classical computing to quantum computing. For example, quantum generalizations of boolean and integer variables were introduced in~\cite{ying2011floyd}. The state space of a quantum boolean variable is the $2$-dimensional Hilbert space $\mathbf{Boolean}=\cH_2$, and the state space of a quantum integer variable is the infinite-dimensional Hilbert space $\mathbf{integer=\cH_{\infty}}$. In \compilername, every kind of quantum variable has its own initialization method and operation. Currently, \compilername contains only finite-dimensional quantum variables, but infinite-dimensional variables will be added in the future. The quantum types used in \compilername are presented in Fig.~\ref{fig:quantum_types_layer}.

The entire quantum types are defined as subclasses of one virtual base class called $\emph{QuantumTypes}\compactT$. The introduction of the virtual base class is only for the purpose of indicating that all of the derived subclasses are quantum objects. From the virtual base class $\emph{QuantumTypes}\compactT$, two extended virtual base classes inherit: $\emph{Vector}\compactT$, which represents a class of quantum variables which share some vector rules, and $\emph{Matrix}\compactT$, which represents a class of quantum operators that share some operator rules.  

Quantum variables come in two basic types:  $\emph{Ket}$ is used to denote a quantum variable of arbitrary dimension, and type $\emph{Bra}$ is the conjugate transpose of $\emph{Ket}$. Two specialized (sub)types $\emph{QBit}$ and $\emph{QBitBra}$ are provided for two-dimension quantum variable. Note that they are compatible when we consider the boolean type as a subtype of an integer. Also, these types must accept a few rules: 
\begin{description}
    \item[Normalized states] For example, a qubit can be written as $\ket{\psi}=\alpha\ket{0}+\beta\ket{1}$. We get either the result $0$ with a probability of $|\alpha|^2$ or the result $1$ with a probability of $|\beta|^2$ when it is measured on a computational basis. Since these probabilities must sum to $1$, it obeys $|\alpha|^2+|\beta|^2=1$. Thus, the length of a vector should be normalized to $1$ during initialization and computation. For convenience, \compilername provides a function $\emph{QBit.NormlizeSelf}()$ to keep the norm of the variable types $\emph{QBit}$ and $\emph{Ket}$.
    \item[Hidden states] It is well-known that the information of a $\emph{QBit}$ or a $\emph{Ket}$ cannot be extracted unless the state is measured. However, as indicated by Nielsen and Chuang in~\cite{Nielsen2000} , ``Nature evolves a closed quantum system of qubits, not performing any `measurements', she apparently does keep track of all the continuous variables describing the state, like $\alpha$ and $\beta$ ". In our platform \compilername, we use the following trick to simulate quantum computing: a quantum state is a black box- each part in the box cooperates with others, but an external viewer knows nothing. Functions and other object methods including unitary transformation or a quantum channel know the quantum state exactly, but a viewer gets nothing about this hidden information unless it is measured. Thus, we classify the state of information storage as a ``$\emph{Protect}$" class, which means that the information of a quantum state cannot be touched easily.
\end{description}

The matrix form is widely used in (the semantics of) the quantum $\cwhile$-language. There are three categorized of matrix: \emph{DensityOperator}$\compactT$, \emph{MeasureMatrixSet} and \emph{SuperOperator}. \emph{DensityOperator}$\compactT$ is also a virtual basic class with two sub-classes: \emph{PureDensityOperator} and \emph{MixDensityOperator}. In fact, the difference between \emph{PureDensityOperator} and \emph{MixDensityOperator} is that only \emph{MixDensityOperator} accepts an ensemble, namely a set of probabilities and their corresponding states, which can be expressed by a \emph{PureDensityOperator}$\compactT$ or a $\emph{Vector}\compactT$. The object quantum variable $\rho$ of $\emph{DensityOperator}\compactT$ satisfies the following two conditions: (1) $\rho$ has trace $1$; (2) $\rho$ is a positive operator. Every operation of objects is set to trigger the verification of these conditions in order to ensure the object is a real density operator. 
$\emph{MeasureMatrixSet}$ is a measurement containing an array of matrix $M=\{M_0,M_1,\ldots,M_n\}$ satisfying a  completeness condition $\sum_i M_i^\dagger M_i=I$. It is very flexible to define a quantum measurement in such a way. Specifically, a plus-minus basis $\{\ket{+},\ket{-}\}$ and a computation basis $\{\ket{0},\ket{1}\}$ are two built-in measurements, and a user can easily use their designed measurement. A $\emph{SuperOperator}$ can be used to simulate an open quantum system.
It uses an array of Kraus operators $\cE=\{E_0,E_1,\ldots,E_n\}$ satisfying $\sum_i E_i^\dagger E_i\leq I$ as a representation. 

\subsubsection{Simulation of quantum behaviors} 
The basis of  simulating quantum computation is to simulate the quantum behaviors defined by the four basic postulates of quantum mechanics~\cite{Nielsen2000}: 
\begin{itemize}
    \item \textbf{Postulate 1}: Associated to any isolated physical system is a complex vector space with an inner product (Hilbert space) known as the state space of the system. The system is completely described by its state vector, which is a unit vector in the system's state space.
    
    {\vskip 4pt}
    
    In the platform \compilername, a function in Math.net called 
    
\ \ \ \ \ \ \ \ \ \ \ \ \ \ \ \emph{double ConjugateDotProduct(Vector$\compactT$ other)} 
    
    is used to support the inner product.
    
    {\vskip 4pt}
    
    \item \textbf{Postulate 2}: The evolution of a closed quantum system is described by a unitary transformation. That is, the state $\ket{\psi}$ of the system at time $t_1$ is related to the state $\ket{\psi^\prime}$ of the system at time $t_2$ by a unitary operator $U$ which depends only on the time $t_1$ and $t_2$. $\ket{\psi^\prime}=U\ket{\psi}$.
    
    {\vskip 4pt}
    
    To simulate this feature in \compilername, we have added the function $UnitaryTrans$ to some of our quantum types such as $\emph{QBit}$, $\emph{Ket}$ and \emph{DensityOperator}$\compactT$ in a closed quantum system. In addition, the static global function $\emph{SuperMatrixTrans}$ is provided to describe the dynamics of an open quantum system as a super-operator $\cE$.  
    
    {\vskip 4pt}
        
    \item \textbf{Postulate 3}:    Quantum measurements are described by a collection $\{M_m\}$ of measurement operators. These are operators acting on the state space of the system being measured. The index $m$ refers to the measurement outcomes that may occur in the experiment. If the state of the quantum system is $\ket{\psi}$  before the measurement, then the probability that the result $m$ occurs is given by 
    $p(m)=\langle\psi|M_m^\dagger M_m|\psi\rangle$ and the state of the system after the measurement is
    $\frac{M_m\ket{\psi}}{\langle\psi|M_m^\dagger M_m |\psi\rangle}$.
    
    {\vskip 4pt}
    
    Quantum measurements are simulated with a modified Monte Carlo method. A detailed description is postponed to the next subsection. 
    
        {\vskip 4pt}
    
    \item \textbf{Postulate 4}: The state space of a composite physical system is the tensor product of the state spaces of the component physical systems. Moreover, if we have systems numbered $1$ through $n$, and system number $i$ is prepared in the state $\ket{\psi}$, then the joint state of the total system is $\ket{\psi_1}\otimes \ket{\psi_2}\otimes \ldots$
    
    {\vskip 4pt}
    
    The function \emph{void KroneckerProduct (Matrix$\compactT$ other, Matrix$\compactT$ result)} is used in the tensor product method, which is embedded in Math.net.
 \end{itemize}

\subsubsection{Simulating measurement with pseudo-random number sampling}

In \compilername, a pseudo-random number sampling method is employed to simulate quantum measurement. It is the numerical experiment generating pseudo-random numbers are distributed according to a given probability distribution~\cite{devroye1986sample}. 

Let a quantum measurement be described by a collection of bounded linear operators $\{M_m\}$ that satisfy a completeness condition $\sum_m M_m^\dagger M_m=I$. $m$ is used to describe the measurement results and the corresponding probability set is denoted as $P$, where $P=\{p_1,p_2,\ldots,p_m\}$. The indexed variable set is denoted as $Y$ where $Y$ can be settled to the value $\{0,1\}$. The current system state is assumed to be the quantum state $\ket{\psi}$, the indexed variables are $Y_1,\ldots,Y_m$ and the probabilities are $\Pr[Y_i=1]=p_i$ where $p_i=\langle\psi|M_i^\dagger M_i|\psi\rangle$, $P=\{p_1,\ldots,p_m\}$. A uniform distribution $X$ from \compilername is used to simulate a random variable $Y$.

$Math.net$ provides a random variable $X$ called  $RandomSource$ which is uniformly distributed on $(0,1)$. Then the interval $[0,1]$ is divided into $m$ intervals as $[0,p_1], (p_1,p_1+p_2], \ldots, (\sum_{i=1}^{m-1}p_i,1]$. The width of interval $i$ equals the probability $p_i$. 

Finally, measurement triggers the strategy in following steps:
\begin{enumerate}
    \item Given a measurement $\{M_m\}$ and the current quantum state $\ket{\psi}$, \compilername computes the set $P=\{p_1,p_2,\ldots,p_m\}$, where $p_i=\langle\psi|M_i^\dagger M_i|\psi\rangle$. This step provides the probability distribution $Y$: $\Pr[Y=i]=p_i$.
    \item \compilername checks the elements of $P$. If there exists any $p_i=0$, discard the index $i$ in the next step. If there exists any $p_i=1$, return the index $i$ as the final result and skip the following steps.
    \item Assuming $P^\prime$ is a set having the same quantity as $P$, \compilername accumulates the distribution $P$ to $P^\prime$ with the rules: for each $p_i$ in $P^\prime$, $p^\prime_i=\sum_i p_i$.
    \item Draw a number $x$ which is a uniformly pseudo-random number distributed between $(0,1)$.
    \item Find $p^\prime_i$, such that $p'_{i-1}\le x$ and $p'_i \geq x$ and return the index $i$. It should be noted that $i=1$ in the case of $x<p'_1$ and $i=m$ in the case of $x>p'_{m-1}$.
\end{enumerate}

The $P$ distribution of the $Y$ variable where $p_i=\Pr(0<Y \leq p_i^\prime)=\sum_i p_i^\prime$ is the simulated distribution using the uniform distribution variable $X$. This method of pseudo-random number sampling was developed for Monte-Carlo simulations and its quality is determined by the quality of the pseudo-number. 

After $i$ is randomly chosen with the distribution $P=\{p_1,\ldots,p_m\}$, the function returns the value of $i$ and the quantum state is modified as an atom operation. According to quantum mechanics, the state $\ket{\psi}$ would be changed into $\ket{\psi^\prime}=\frac{M_i\ket{\psi}}{\sqrt{\langle\psi|M_i^\dagger M_i|\psi\rangle}}$.

\subsubsection{Simulating operational semantics of quantum \textbf{while}-language}

Simulating the computation of a program written in the quantum \textbf{while}-language is based on simulating the operational semantics of the language. To clearly delineate coding in mixed classic-quantum programs, quantum $\mathbf{if}$-clause are denoted as $\mathbf{cif}$ and quantum $\mathbf{while}$-clause are denoted as $\mathbf{cwhile}$ in quantum simulation engine. To simulate these two functions, the related function methods are encapsulated in Quantum Mechanics Library.

The execution of a quantum program can be conveniently described in terms of transitions between configurations.
\begin{definition}
    A quantum configuration is a pair $\langle S,\rho \rangle$, where:
    \begin{itemize}
        \item $S$ is a quantum program or the empty program $E$ (termination);
        \item $\rho$ is a partial density operator used to indicate the (global) state of quantum variables.
    \end{itemize}
\end{definition}

With the preparations in the previous subsections, we are able to simulate the transition rules that define the operational semantics of the quantum while-language: 

\begin{description}
    \item[Skip] 
    
    $$\frac{}{\langle \mathbf{skip},\rho\rangle \to \langle \mathbf{E},\rho\rangle}\,.$$

    The statement $\mathbf{skip}$ does nothing and terminates immediately. Both $I$-identity operation and the null clause satisfy this procedure requirement for simulation in \compilername. 
    
    {\vskip 4pt}
    
    \item[Initialization] 
    
    $$\frac{}{\langle q:=\ket{0},\rho\rangle \to \langle \mathbf{E},\rho_0^q \rangle}\,,$$  where 
    \[
    \rho_0^q=
    \left\{
    \begin{array}{lll}
    \ket{0}_q\bra{0}\rho\ket{0}_q\bra{0}&+&\ket{0}_q\bra{1}\rho\ket{1}_q\bra{0}\ \textrm{if}\ \,type(q)=Boolean,\\
    \\
    \sum_{n=-\infty}^{\infty}\ket{0}_q\bra{n}&\rho&\ket{n}_q\bra{0}\ \textrm{if}\ \, type(q)=Integer .
    \end{array}
    \right.
    \]
    The initialization statement ``$q:=\ket{0}$" sets the quantum variable $q$ to the basis state $\ket{0}$.  
    
    {\vskip 4pt}
    
    In \compilername, initialization has two forms. When the variable $q$ is a $\emph{QBit}$, it is explained as $\llbracket q:=\ket{0}\rrbracket (\rho)=\ket{0}\bra{0}\rho\ket{0}\bra{0}+\ket{0}\bra{1}\rho\ket{1}\bra{0} $; otherwise, it is explained as $\llrr{q:=\ket{0}}(\rho)=\sum_{n=0}^{d}\ket{0}\bra{n}\rho\ket{n}\bra{0}$, where $d$ is the dimension of the quantum variable $q$. Moreover, a more flexible initialization method is provided with the help of unitary transformation.
    
    {\vskip 4pt}
    
    \item[Unitary revolution] 
    $$\frac{}{\langle \bar{q}:=U[\bar{q}],\rho\rangle \to \langle \mathbf{E},U\rho U^\dagger\rangle}\,.$$
    
    The statement  ``$\bar{q}:=U[\bar{q}]$" means that the unitary gate $U$ is performed on the quantum register $\bar{q}$ leaving other variables unchanged. 
    
    {\vskip 4pt}
    
 A corresponding method called \emph{QuantumTypes$\compactT$} \emph{.UnitaryTrans(Matrix$\compactT$\,other)} has been designed for $\emph{QBit}, \emph{Ket}, \emph{DensityOperator}\compactT$ objects to perform this function. This function accepts a unitary operator and performs the operator on the variable with null returns. We have also provided a global function called 
    
    \ \ \ \ \ \ \ \ \ \ \ \ \emph{UnitaryGlobalTrans(QuantumType$\compactT$,   Matrix$\compactT$)} 
    
    that perform an arbitrary unitary matrix on quantum variables.
    
    {\vskip 4pt}
    
    The quantum $\cwhile$-languages do not include any assignment claim for a pure state because a unitary operator $U$ exists for any pure state $\ket{\psi}$ satisfies $\ket{\psi}=U\ket{0}$. Therefore, any pure state can be produced from a combination of an initialization clause and a unitary transformation clause. However, for convenience, \compilername provides a flexible state claim to initialize a $\emph{QBit}$, or a $\emph{Ket}$ using a vector, and to initialize a $\emph{DensityOperator}\compactT$ using a positive matrix.

    \item[Sequential composition]
    
    $$\frac{\langle S_1,\rho\rangle \to \langle S_1',\rho\rangle}{\langle S_1;S_2,\rho\rangle \to \langle S_1';S_2,\rho'\rangle}\,.$$
        
    The current version of the quantum $\cwhile$-language is not designed for concurrent programming. Thus sequential composition is spontaneous.
    
    {\vskip 4pt}
    
    \item[Case statement]    $$\frac{}{\langle \mathbf{if}(\square m\cdot M[\bar{q}]=m\to S_m)\mathbf{fi},\rho \rangle \to \langle S_m,M_m\rho M_m^\dagger\rangle}\,,$$ for each possible outcome $m$ of measurement $M=\{M_m\} \,.$    
    
    {\vskip 4pt}
    
    The first step in executing of the case statement is performing a measurement $M$ on the quantum variable $\bar{q}$ and observing the output result index. The corresponding subprogram $S_m$ is then chosen according to the index.
    
        {\vskip 4pt}
    
    Case statements in \compilername use an encapsulated function with the prototype \[\mathbf{cif}(QuantumTypes\compactT,MeasureMatrixSet,Func\compactT,Func\compactT\ldots)\].
    
    By default, the  $Func\compactT$ sequence is a subprogram corresponding to a measurement output index, i.e., the $n$th $Func\compactT$ corresponds to the $n$th measurement output index.
    We have also considered cases where the user has not provided a subprogram corresponding to every measurement output index. In these situations, \compilername's strategy is to automatically skip that clause if the outcome index exceeds $Func\compactT$ number. In fact, nothing to be done on variables excepted a measurement in this case.
    
    Another difference between a  classical and a quantum case statement is that quantum case statement variables must be modified into the state corresponding the measurement output index after performing a measurement. We call the function 
    \[int\,Measu2ResultIndex(MeasureMatrixSet)\] to return the measurement result and go to the correct subprogram, then it would call the $void\, StateChange(int)$ inherently, which changes the variable $\bar{q}$ to the corresponding state after the measurement. 
    
    {\vskip 4pt}
    
    \item[Loop statement] 
    
    $$({\rm L0})\ \ \frac{}{\langle \mathbf{while}(M[\bar{q}]=1) \mathbf{do}\, S \, \mathbf{od},\rho \rangle \to \langle \mathbf{E},M_0\rho M_0^\dagger\rangle}\,,\ \ \ $$
    $$({\rm L1})\ \ \frac{}{\langle \mathbf{while}(M[\bar{q}]=1) \mathbf{do}\, S \, \mathbf{od},\rho \rangle \to \langle S;\mathbf{while}(M[\bar{q}]=1) \mathbf{do}\, S \, \mathbf{od},M_1\rho M_1^\dagger\rangle}\,.$$

    To implement this loop statement in \compilername, we use an encapsulated function with the prototype     \[\mathbf{cwhile}(QuantumTypes\compactT, MeasureMatrixSet,int,Func\compactT)\]
    This function accepts quantum types, a measurement, and an integer. Then, it compares the measurement result with the given integer in the guard. If the guard has a value of `$1$', it enters into the loop body, otherwise it terminates. In addition, the state will have been changed after being measured in the guard. The function 
    \[int\, Measu2ResultIndex(MeasureMatrixSet)\] is called to return the guard index and go to the correct subprogram, then it calls the $void\, StateChange(int)$ inherently as per the case statement.   
\end{description}

\section{Experiments}

Here, we present three experiments to show the power of our quantum programming environment: Qloop, BB84 and Grover's search algorithm. Readers can find more details in the Appendices.
\begin{description}
    \item[Qloop] Qloop case is a ``Hello world" example that includes a quantum channel, a quantum measurement, a quantum $\cwhile$-clause and some quantum variables. Basically, it can be regarded as a simplified quantum walk. This test illustrates three main features of the \compilername platform, super-operators, unitary transformations and quantum measurement.
        
            {\vskip 4pt}
            
    The basic idea of a Qloop is to perform a super-operator on a quantum state and leave the state changed. A counter is used to record the number of times the state enters into different branches. A measurement is taken in every shots and the counter should show the predicted probability for the state.
    
        {\vskip 4pt}

    \item[BB84] BB84 is a quantum key distribution (QKD) protocol developed by Bennett and Brassard in 1984~\cite{bennett2014quantum}. The protocol is an already-proven security protocol~\cite{shor2000simple} that relies on the no-cloning theorem. Using this protocol Alice and Bob reach an agreement about a classical key string that can be used to encrypt classical bits.
    
        {\vskip 4pt}
    
    Several different scenarios are considered in this experiment. The simple BB84 case outlines the basic communication procedure between two clients: Alice and Bob. The multi-client BB84 case illustrates a more practical example where one Alice generates the raw keys, and many Bobs make an agreement key with Alice. The most interesting case is the BB84 protocol in a channel with quantum noise.  Because no real quantum systems are ever perfectly closed, super-operators can serve as a key tool for describing the dynamics of open quantum systems. 
    In this case, influencing factors for QKD are explored. The package length and sampling percentages are crucial to real QKD protocol under quantum noise. With \compilername, different parameters are tested in different channels which can be adjusted for practical use of this protocol.  
    
        {\vskip 4pt}
        
    \item[Grover's search algorithm] Grover's search algorithm is an impressive algorithm in the quantum domain.
    It solves search task in disorderly databases consisting of $N$ elements, indexed by number $0,1,\ldots,N-1$ with an oracle. The oracle returns it answers according to position and can find solutions with a high probability within $O(1/N)$ error and $O(\sqrt{N})$ steps. 
    
    A more general multi-object Grover's search is also considered that supposes there is more than one answer (position) for the oracle to find. In this case, we use a blind box strategy that reverses the proper position of the answer. This experiment reveals that Grover's algorithm leads to an avalanche of errors in a multi-object setting, indicating that the algorithm needs be modified in some way.
\end{description} 

\section{Conclusions}

This paper presents a new software platform, \compilername, for programming quantum computers. \compilername includes an embedded quantum $\cwhile$-language, a quantum simulator, and quantum program analysis and verification toolkits. The platform can be used to simulate quantum algorithms, analyze the termination and average running time of quantum programs, and verify program correctness.

Throughout the paper, we demonstrate how to use \compilername to simulate quantum behaviors on classical platforms using a combination of components. We discuss simulating measurement with pseudo-random number sampling, and how to generate the syntax and semantics of the quantum $\cwhile$-language.

Active development of \compilername is ongoing. The tensor product is a clumsy way of emulating quantum circuits. We may need to consider timing and entanglement analysis inspired by~\cite{scaffCC} to extend \compilername's quantum computing power. The Termination and Average Running Time modules need to be unified into one format for syntax, and we are considering how to split classical and quantum coding for verification purposes.

Interfaces for different quantum computation programs, such as \liquid, ScaffCC and even the real quantum computation platform from IBM's quantum experiment also need to be considered. These diversified platforms often can provide different views of one quantum program.

\section*{Acknowledgments}
% \noindent\rule{8.6cm}{2pt}
 We were grateful to Michael Blumenstein, Ian Burnett, Yuan Feng, and Glenn Wightwick for their helpful discussions and support to this project.

%\subsubsection*{Acknowledgments.} Thank 

%\begin{thebibliography}{4}
%\end{thebibliography}
\bibliographystyle{plain}       % APS-like style for physics
\bibliography{ref}   % name your BibTeX data base
%\bibliography{D:/Users/klinu/OneDrive/StudyNote/ref} 

\appendix
\section{Setup and configuration of \compilername}
\compilername mainly relies on IDE (Visual Studio) to provide the details of the program. After completing a program using \compilername, the programmer needs to build and compile it. This feature is considered to be an essential component because a smarter IDE is a basic way of ensuring the syntax is correct as programs grow in size and complexity. This feature is unlike IScasMC or QPAT which are not able to execute a program.

\emph{NuGet} is a part of the .Net development platform and it is used in \compilername to manage the packages. All packages used to provide functions, such as matrix computation, random number generation, and Roslyn, etc., can be automatically controlled by \emph{NuGet}. To add all the essential packages, a user needs only add the NuGet feed v3 ``https://api.nuget.org/v3/index.json" to the Visual Studio 2017 configuration. This will provide the package resources and automatically configure them for the platform.

\compilername is compatible with any version of Visual Studio 2015 and later. However, we recommend the Enterprise version of Visual Studio 2017 because of some of its premium features, such as the ability to draw quantum circuits with the DGML tools, the most up-to-date Math.net, etc. Examples are stored in the sub-folder UnitTest. All entry-level examples can be found in the `Program.cs' file in UnitTest. 

\section{Experiment-Qloop case}
\begin{figure}[!htp]  
    \centering  
    \includegraphics[width=70mm]{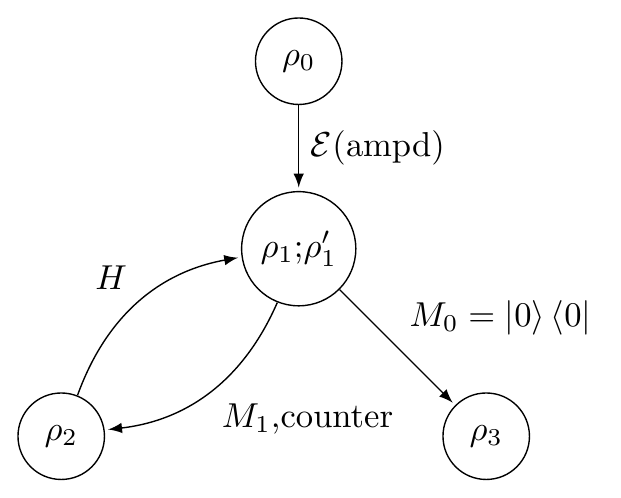}  
    \caption{Qloop}  
    \label{fig:qloop}  
\end{figure}

The first example showcases the Qloop case. It uses quantum channels, measurement, quantum $\cwhile$-clause and quantum variables. The Qloop case can also be treated as a simplified quantum walk. The flow path is shown in Fig.~\ref{fig:qloop}.

\subsection{Input and output}
%The definition is as below:
{Input}:
\begin{itemize}
    \item $\rho_0:=\ket{+}\bra{+}$;
    \item $\cE:=\{ E_0=\ket{0}\bra{0}+\ket{1}\bra{1}/\sqrt{2},E_1=\ket{0}\bra{1}/\sqrt{2}\}$;
    \item $M:=\{M_0=\ket{0}\bra{0},M_1=\ket{1}\bra{1}\}$;
    \item $H:=\ket{+}\bra{0}+\ket{-}\bra{1}$;
    \item $Counter:=0$.
\end{itemize}

Output:
\begin{itemize}
    \item $num$: the number of circles is $num$.
\end{itemize}

\subsection{Results}
\begin{figure}[!htp]  
    \centering  
    \includegraphics[width=90mm]{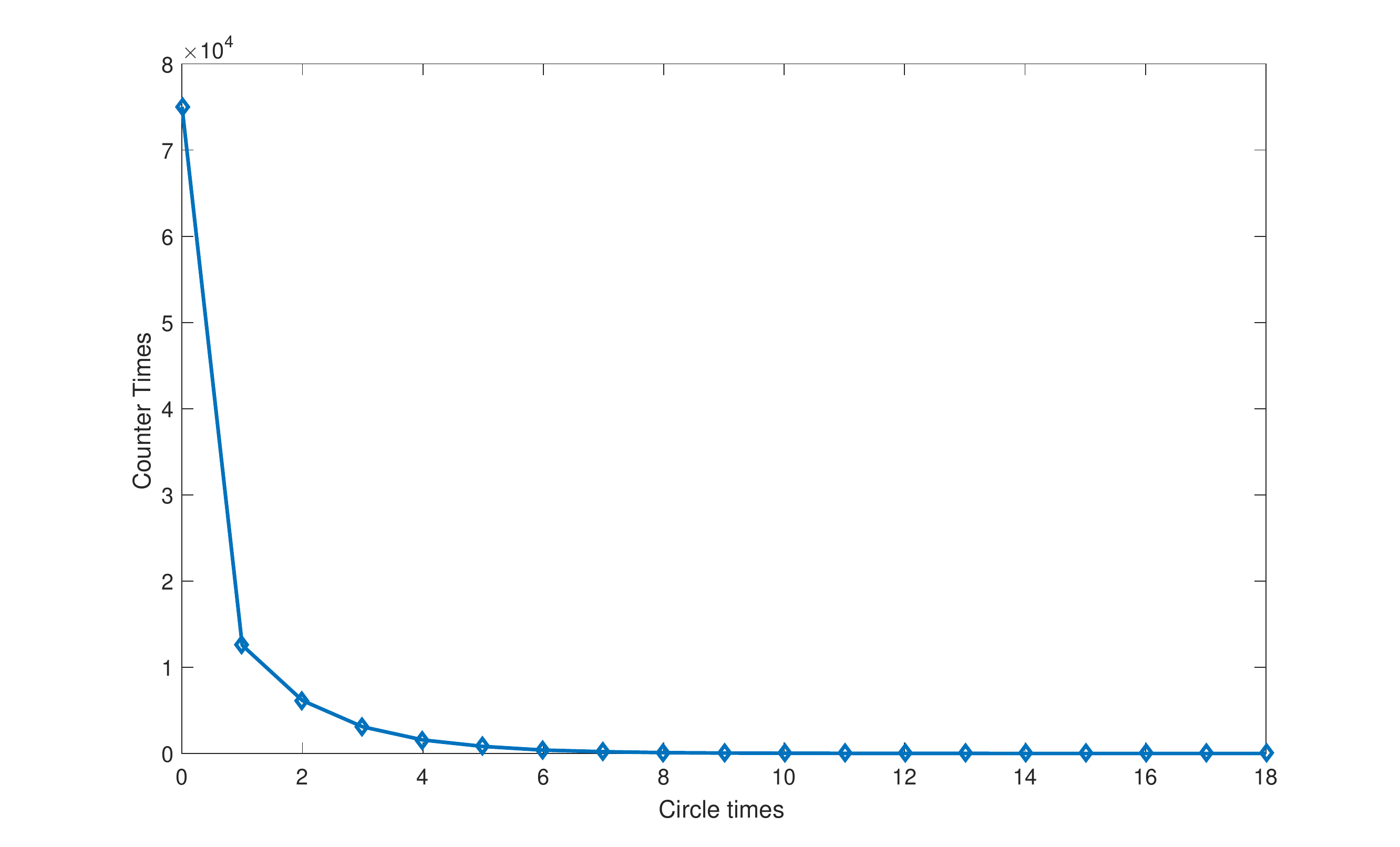}  
    \caption{Qloop data}  
    \label{fig:qloop_data}  
\end{figure}

The Qloop experiment is executed about 100000 shots and the results are shown in the Fig.~\ref{fig:qloop_data}.

\subsection{Features and analysis}
After calculation, it is clear that $\rho_1=\cE(\rho_0)=\frac{3}{4}\op{0}{0}+\frac{1}{4}\op{1}{1}+\frac{1}{2\sqrt{2}}\op{0}{1}+\frac{1}{2\sqrt{2}}\op{1}{0}$, $\rho_2=\op{1}{1}$, $\rho_1^\prime=\op{+}{+}$ and $\rho_3=\op{0}{0}$.\\

The three main features of this experiment include super-operators, unitary transformation, and measurement operation. In addition, processes that consider a qubit's collapse and measurement probability are involved as part of quantum mechanics.
\begin{itemize}
    \item Super-operator operation. The initial state passes through a quantum channel and becomes $\rho_1$. Let $M$ be performed on the state $\rho_1$ in each shot. There is a $\frac{3}{4}$ probability that the state would change to $\rho_3$ and then terminate. Likewise, there is a $\frac{1}{4}$ probability of moving in a circle and having the process recorded by the counter. So if the program is executed many times, such as in a $100000$ shot experiment, the counter should show the state enters the circle about $25000$ times.
    \item Measurement operation and unitary transformation. After the first measurement, $\rho_1$ would change to $\rho_2$ and continue or it would change to $\rho_3$ and terminates. If the state becomes $\rho_2$, after a Hadamard operator which is a unitary transformation, it becomes $\rho_1^\prime=\op{+}{+}$ and counter records the circle once. When a measurement $M$ is performed on the state, we can assert that almost half the time  $\rho_1^\prime$ becomes $\op{0}{0}$ and half the time it becomes $\op{1}{1}$. If the result is $\op{1}{1}$, it will enter into the loop body again and is recorded by our counter. In total, the counter number shows how many circles the state enters into. Obviously, this decreases at almost half the rate of a geometric progression, as in say $1-12556$, $2-6140$, $3-3095$, $\ldots$  
\end{itemize}

\section{BB84 case}
BB84 is a basic quantum key distribution (QKD) protocol developed by Bennett and Brassard in 1984~\cite{bennett2014quantum}. 
\subsection{Simple BB84 case}
\label{sec:simpBB84}
\begin{figure}[!htp]  
    \centering  
    \includegraphics[width=170mm]{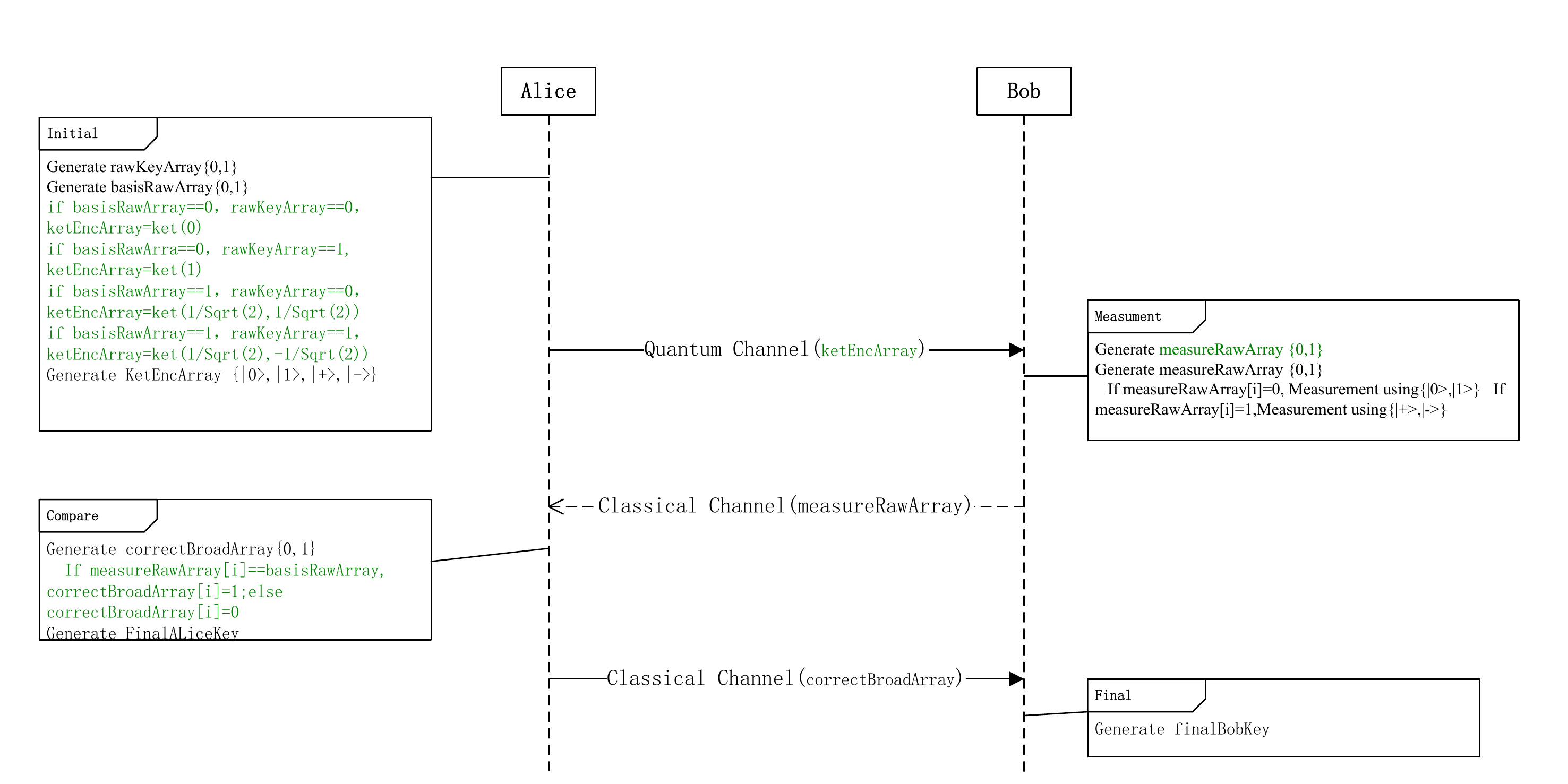}  
    \caption{Simple BB84 protocol}  
    \label{fig:singleBB84}  
\end{figure}

In this case, a client-server model is used as a prototype for a multi-user communication protocol. A ``quantum type converter" is used to convert a `$Ket$' into a density operator. For simplicity and clarity, this example only  `$Ket$' quantum types is considered, not quantum channels or Eves.  The flow path is shown in Fig.~\ref{fig:singleBB84}.

The entire flow path follows,
\begin{enumerate}
    \item Alice randomly generates a sequence of classical bits called a $\emph{rawKeyArray}$. Candidates from this raw key sequence are chosen to construct the final agreement key. The sequence length is determined by user input.
    \item Alice also randomly generates a sequence of classical bits called $\emph{basisRawArray}$. This sequence indicates the chosen basis to be used in next step. Alice and Bob share a rule before the protocol:
    \begin{itemize}
        \item They use $\{\ket{+},\ket{-}\}$ or $\{\ket{0},\ket{1} \}$ to encode the information.
        \item A classical bit 0 indicates a $\{\ket{0},\ket{1} \}$ basis while a classical bit 1 indicates $\{\ket{+},\ket{-} \}$. This rule is used to generate Alice's  qubits and to check Bob's basis.
    \end{itemize}
    \item Alice generates a sequence of quantum bits called $\emph{KetEncArray}$, one by one according to the rules below,
    \begin{itemize}
        \item If the $\emph{basisRawArray[i]}$ in position [i] is $0$ and the $\emph{rawKeyArray[i]}$ in position [i] is $0$, $\emph{KetEncArray[i]}$ would be $\ket{0}$.
        \item If the $\emph{basisRawArray[i]}$ in position [i] is $0$ and the $\emph{rawKeyArray[i]}$ in position [i] is $1$, $\emph{KetEncArray[i]}$ would be $\ket{1}$.
        \item If the $\emph{basisRawArray[i]}$ in position [i] is $1$ and the $\emph{rawKeyArray[i]}$ in position [i] is $0$, $\emph{KetEncArray[i]}$ would be $\ket{+}$.
        \item If the $\emph{basisRawArray[i]}$ in position [i] is $1$ and the $\emph{rawKeyArray[i]}$ in position [i] is $1$, $\emph{KetEncArray[i]}$ would be $\ket{-}$.
    \end{itemize} 
    \item Alice sends the $\emph{KetEncArray}$ through a quantum channel. In this case, she sends it through the $I$ channel. 
    \item Bob receives the $\emph{KetEncArray}$ through the quantum channel.
    \item Bob randomly generates a sequence of classical bits called $\emph{measureRawArray}$. This sequence indicates the chosen basis to be used in next step. 
    \item Bob generates a sequence of classical bits called $\emph{tempResult}$, using quantum measurement according to the rules:
    \begin{itemize}
        \item If the $\emph{measureRawArray[i]}$ in [i] position is a classical bit 0, Bob uses a $\{\ket{0},\ket{1} \}$ basis to measure the $\emph{KetEncArray[i]}$ while a classical bit 1 indicates using a $\{\ket{+},\ket{-} \}$ basis.
    \end{itemize}
    \item Bob broadcasts the $\emph{measureRawArray}$ to Alice using a classical channel.
    \item Alice generates a sequence of classical bits called $\emph{correctBroadArray}$, by comparing Bob's basis $\emph{measureRawArray}$ and her basis $\emph{basisRawArray}$. If the position [i] is correct, the $\emph{correctBroadArray[i]}$ would be $1$, otherwise would be $0$.
    \item Alice sends the sequence $\emph{correctBroadArray}$ to Bob.
    \item Alice generates a sequence of classical bits called $FinalALiceKey$ using the rule:
    \begin{itemize}
        \item If position [i] in $\emph{correctBroadArray[i]}$ is 1, she keeps $\emph{rawKeyArray[i]}$ and copies it to $\emph{FinalALiceKey}$ , else she discards $\emph{rawKeyArray[i]}$.
    \end{itemize}
    \item Bob generates a sequence of classical bits called $FinalBobKey$ using the rule:
    \begin{itemize}
        \item If position [i] in $\emph{correctBroadArray[i]}$ is 1, he keeps $\emph{tempResult[i]}$ and copies it to $\emph{FinalBobKey[i]}$, else he discards $\emph{tempResult[i]}$.
    \end{itemize}
    \item GlobalView: We use a function compare whether every position [i] in $\emph{FinalALiceKey}$ and $\emph{FinalBobKey[i]}$ are the same.
\end{enumerate}

This case shows some useful features,
\begin{itemize}
    \item Client-server mode. The process uses a client-server model to simulate the BB84 protocol. The model includes many implicit features, such as waiting threads and concurrent communications which are also used in the next example.
    \item Measurement. According to theory, choosing a random measurement basis may arrive at half of the correct result. As a result, the agreement of classical shared bits should be almost half the length of the raw keys.    
\end{itemize}

\subsection{BB84 case, multi-client}
The multi-client BB84 model offers a more attractive and practical example. In this model, there is one Alice to generate the raw keys and many Bobs to construct an agreement key with Alice. 

In this case, users can specify the number of clients. Also, a typical BB84 flow path would occur for every client-server pair of this model. 

\begin{itemize}
    \item Threads Model. Many clients are generated and communicate with Alice. Each of them finally reaches an agreement.
    \item Measurement threads. In this case, Alice generates raw keys, and Bob measures the quantum bits. However, this raises a serious question that when a client is considered to generate a raw key while the server measure, how can we ensure the server correctly and fairly conducts the measurement for the server.
\end{itemize}

\subsection{BB84 case with noise}
A practical topic for the \compilername is to consider the BB84 model with noisy quantum channels. Noisy quantum operations are the key tools for the description of the dynamics of open quantum systems. 

In this example, different channels such as bit flip, depolarizing, amplitude damping and $I$-identity channels are described by quantum operations performing as the evolution of quantum systems in a wide variety of circumstances. Alice and Bob use these quantum channels to communicate with each other via the BB84 protocol as Fig.~\ref{fig:singleBB84} shows. During communication, verification steps also need to be considered.

\subsubsection{Input and output}
In this example, the basic quantum channels are defined as follows: 
%depolarizing

deplarizing channel with noise parameter $p=0.5$,
\[
\cE:=\{
\begin{bmatrix}
\frac{\sqrt{5}}{\sqrt{8}}&0\\
0&\frac{\sqrt{5}}{\sqrt{8}}\\
\end{bmatrix},
\begin{bmatrix}
0&\frac{1}{\sqrt{8}}\\
\frac{1}{\sqrt{8}}&0\\
\end{bmatrix}
\begin{bmatrix}
0&\frac{-i}{\sqrt{8}}\\
\frac{i}{\sqrt{8}}&0\\
\end{bmatrix}
\begin{bmatrix}
\frac{1}{\sqrt{8}}&0\\
0&-\frac{1}{\sqrt{8}}\\
\end{bmatrix}
\}
\,;
\]

%amplitude
amplitude damping channel with noise parameter $\gamma=0.5$,
\[
\cE:=\{
\begin{bmatrix}
1&0\\
0&\frac{1}{\sqrt{2}}\\
\end{bmatrix},
\begin{bmatrix}
0&\frac{1}{\sqrt{2}}\\
0&0\\
\end{bmatrix}
\}
\,;
\]

%And three kinds of bit flip channel:

%bit flip
bit flip channel with noise parameter $p=0.25$,
\[
\cE:=\{
\begin{bmatrix}
\frac{1}{2}&0\\
0&\frac{1}{2}\\
\end{bmatrix},
\begin{bmatrix}
0&\frac{\sqrt{3}}{2}\\
\frac{\sqrt{3}}{2}&0\\
\end{bmatrix}
\}
\,;
\]

bit flip channel with noise parameter $p=0.5$,
\[
\cE:=\{
\begin{bmatrix}
\frac{1}{\sqrt{2}}&0\\
0&\frac{1}{\sqrt{2}}\\
\end{bmatrix},
\begin{bmatrix}
0&\frac{1}{\sqrt{2}}\\
\frac{1}{\sqrt{2}}&0\\
\end{bmatrix}
\}
\,;
\]

bit flip channel with noise parameter $p=0.75$,
\[
\cE:=\{
\begin{bmatrix}
\frac{\sqrt{3}}{2}&0\\
0&\frac{\sqrt{3}}{2}\\
\end{bmatrix},
\begin{bmatrix}
0&\frac{1}{2}\\
\frac{1}{2}&0\\
\end{bmatrix}
\}
\,.
\]

The flow path follows the simple BB84 protocol shown in Fig.~\ref{fig:singleBB84}. The only differences are in Step 4 and a sampling step is added.
\begin{itemize}
    \item  Alice sends the $\emph{KetEncArray}$ through a quantum channel. In this case, it is one of the channels mentioned above.
    \item Sampling check step: Alice randomly publishes some sampling positions randomly with the bits against these positions in her own key string. Bob checks these bits against his own key strings. If all the bits in these sampling strings are the same, he believes the key distribution is a success; Otherwise, the connection fails.  
\end{itemize} 

For a statistical quantity characterizing success in a channel with the BB84 protocol, we executed a 100-shot experiment for each channel. In every shot for every channel, different sampling percentages and package lengths were considered. The tables and figures are provided in Fig.~\ref{fig:BB84channel} that show the trade-off between success times, different sampling proportions and package lengths in each of the quantum channels.

\subsubsection{Results}
\begin{figure}[!h]
    %\begin{tabular}{cc}
    \begin{minipage}{0.38\linewidth}
        \centerline{\includegraphics[width=220px]{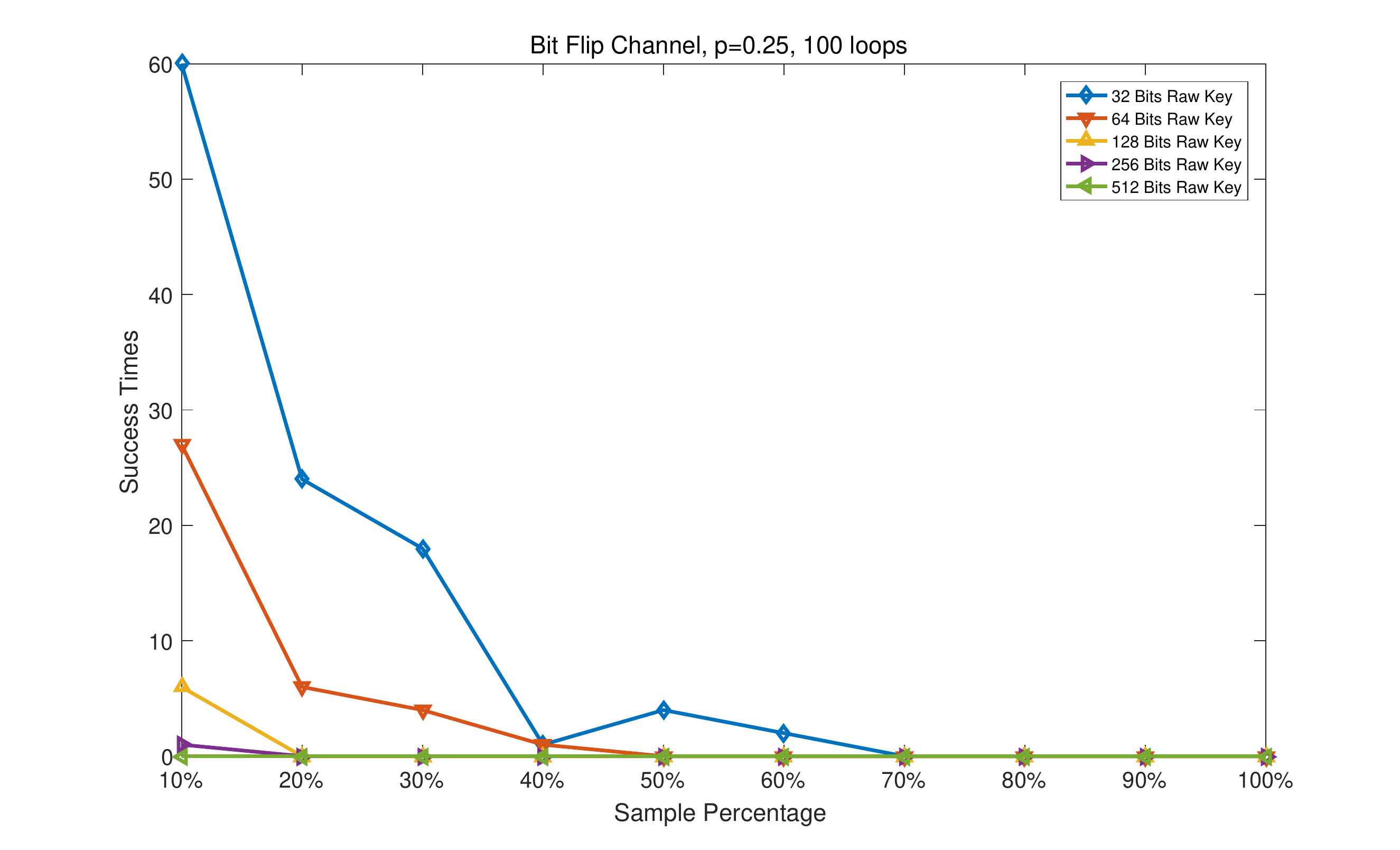}}
        \centerline{(a) Bit Flip Channel, $p=0.25$, $loops=100$}
    \end{minipage}
    \hfill
    \begin{minipage}{0.38\linewidth}
        \centerline{\includegraphics[width=220px]{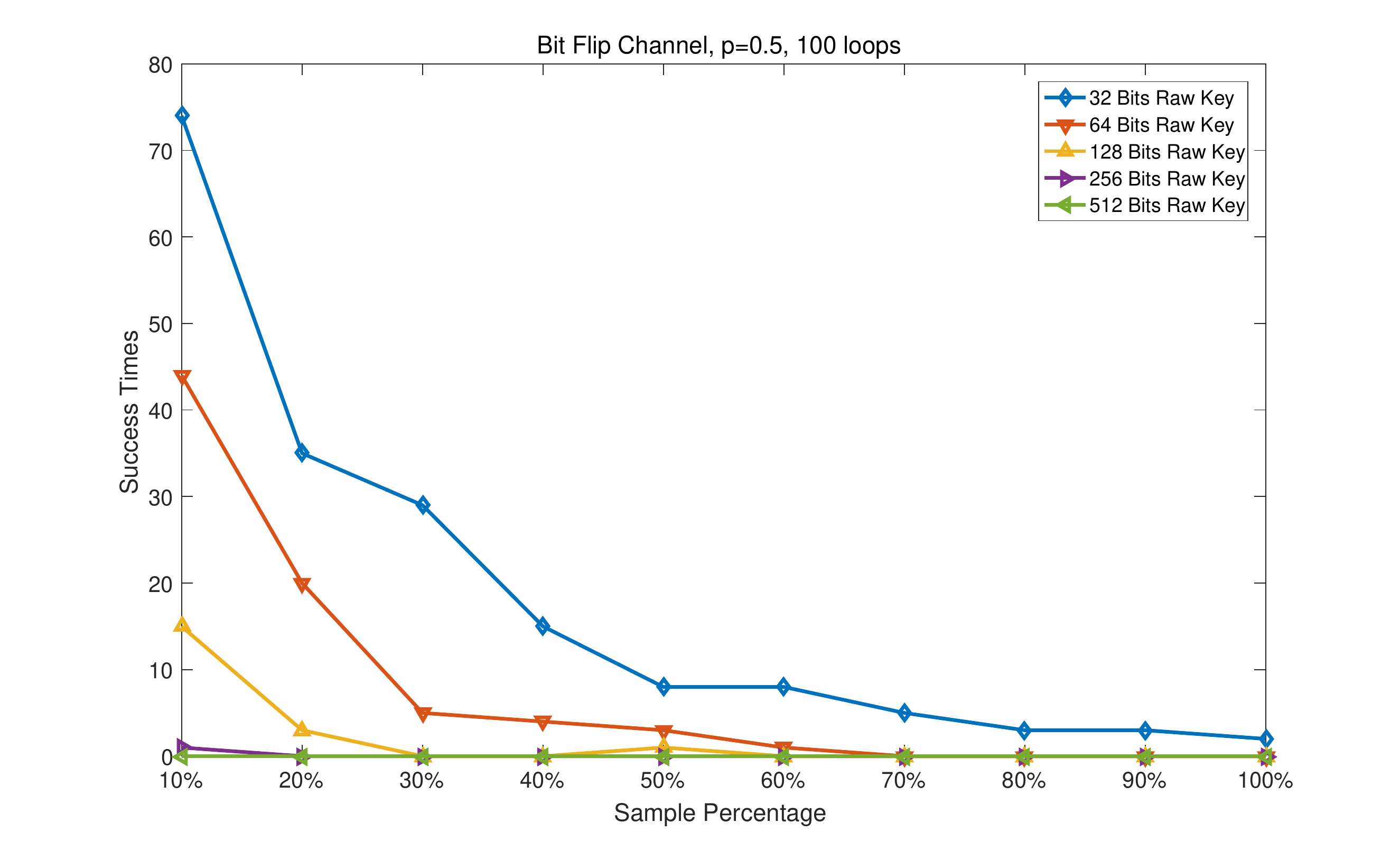}}
        \centerline{(b) Bit Flip Channel, $p=0.5$, $loops=100$}
    \end{minipage}
    \vfill
    \begin{minipage}{0.38\linewidth}
        \centerline{\includegraphics[width=220px]{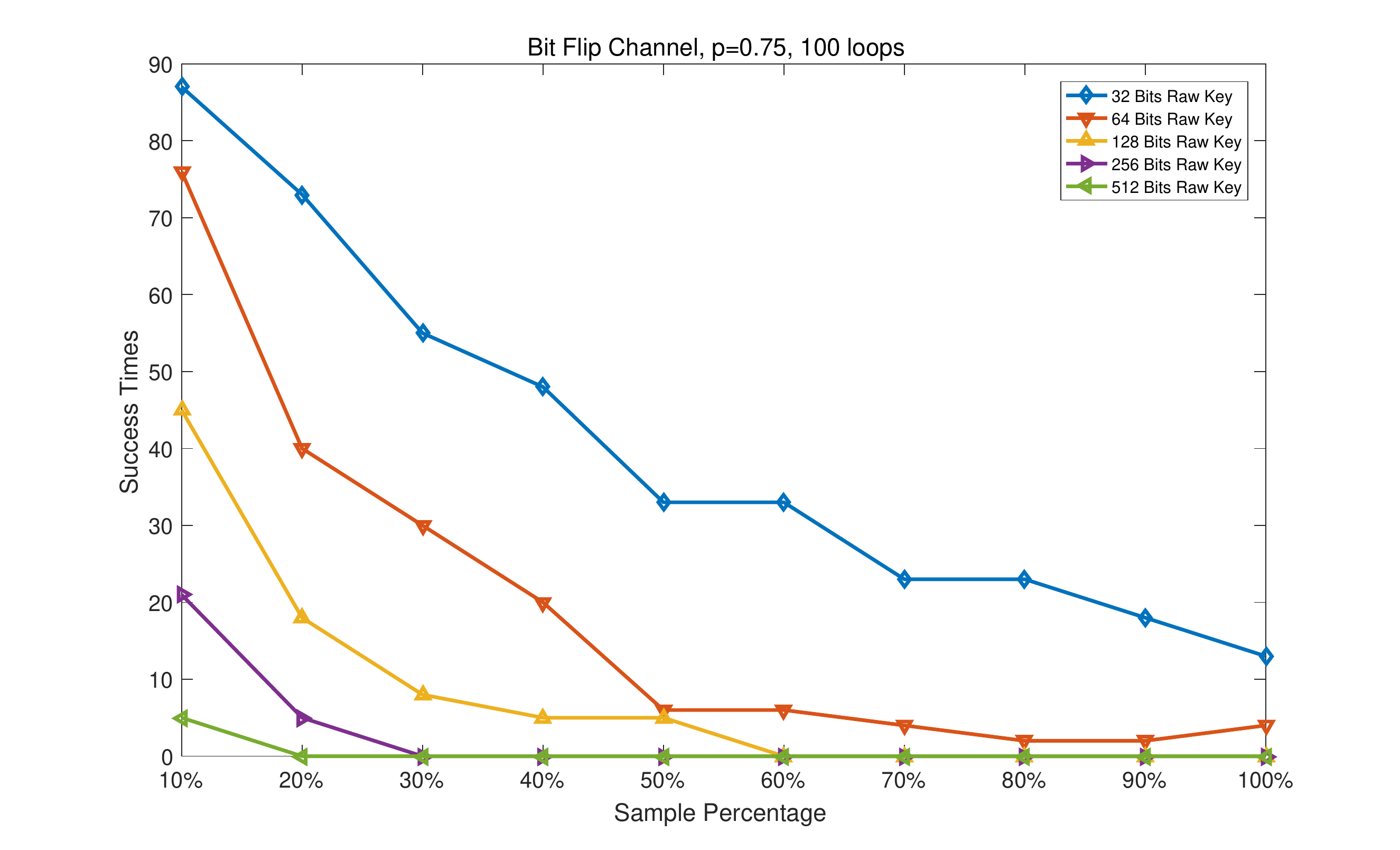}}
        \centerline{(c) Bit Flip Channel, $p=0.75$, $loops=100$}
    \end{minipage}
    \hfill
    \begin{minipage}{0.38\linewidth}
        \centerline{\includegraphics[width=220px]{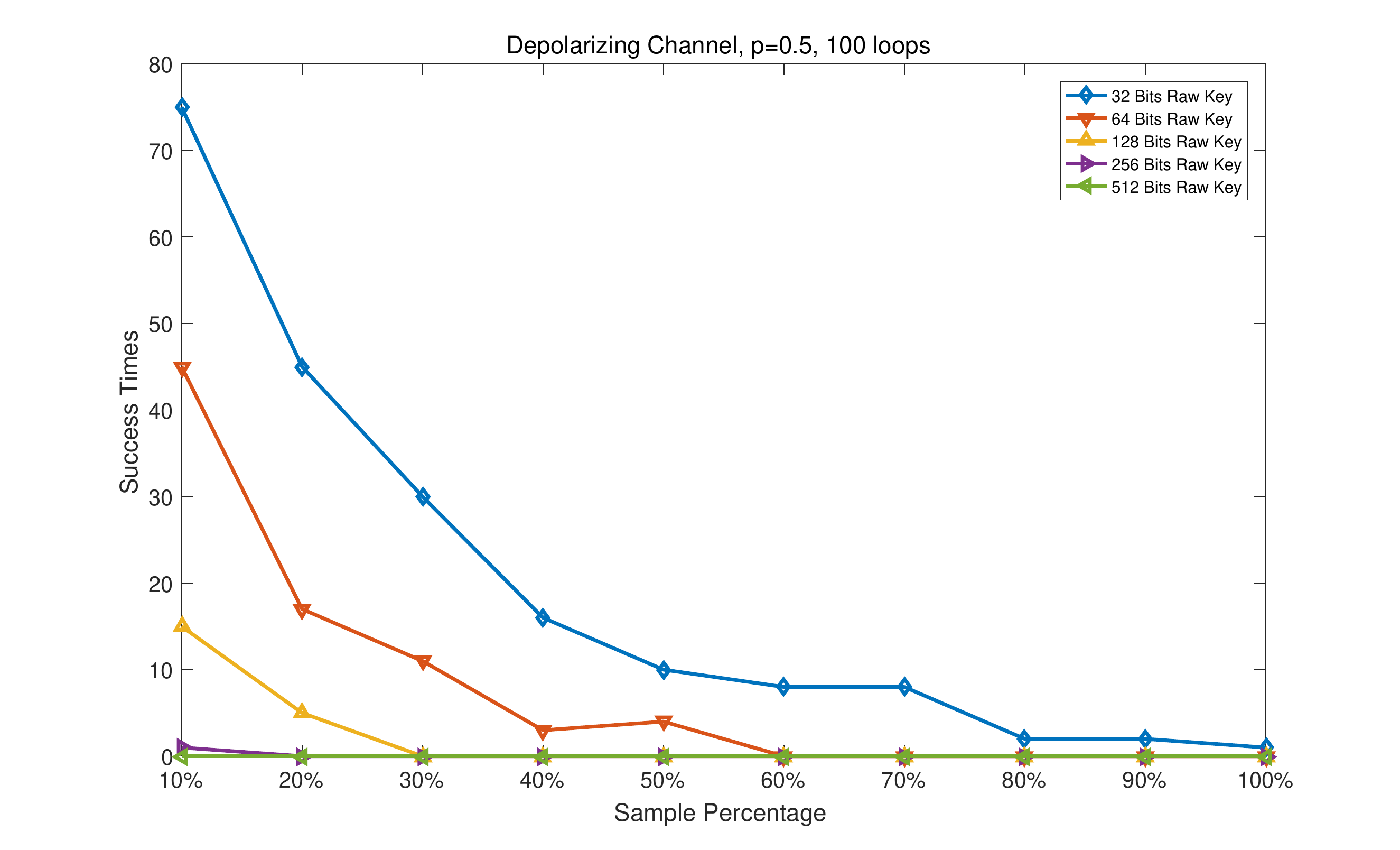}}
        \centerline{(d) Depolarizing Channel, $p=0.5$, $loops=100$}
    \end{minipage}
    \vfill
    \begin{minipage}{0.38\linewidth}
        \centerline{\includegraphics[width=220px]{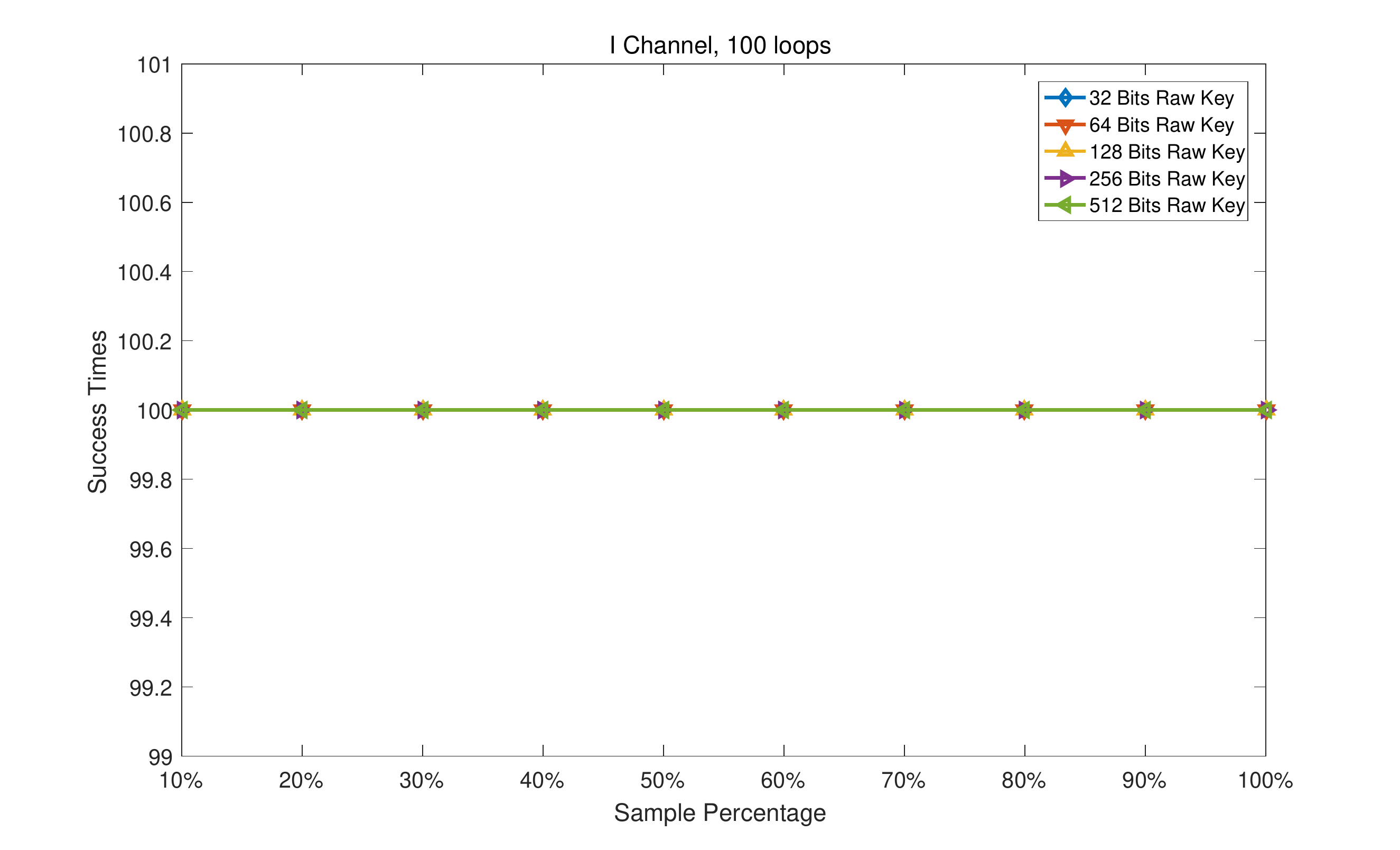}}
        \centerline{(e) $I$-channel, $loops=100$}
    \end{minipage}
    \hfill
    \begin{minipage}{0.38\linewidth}    
        \centerline{\includegraphics[width=220px]{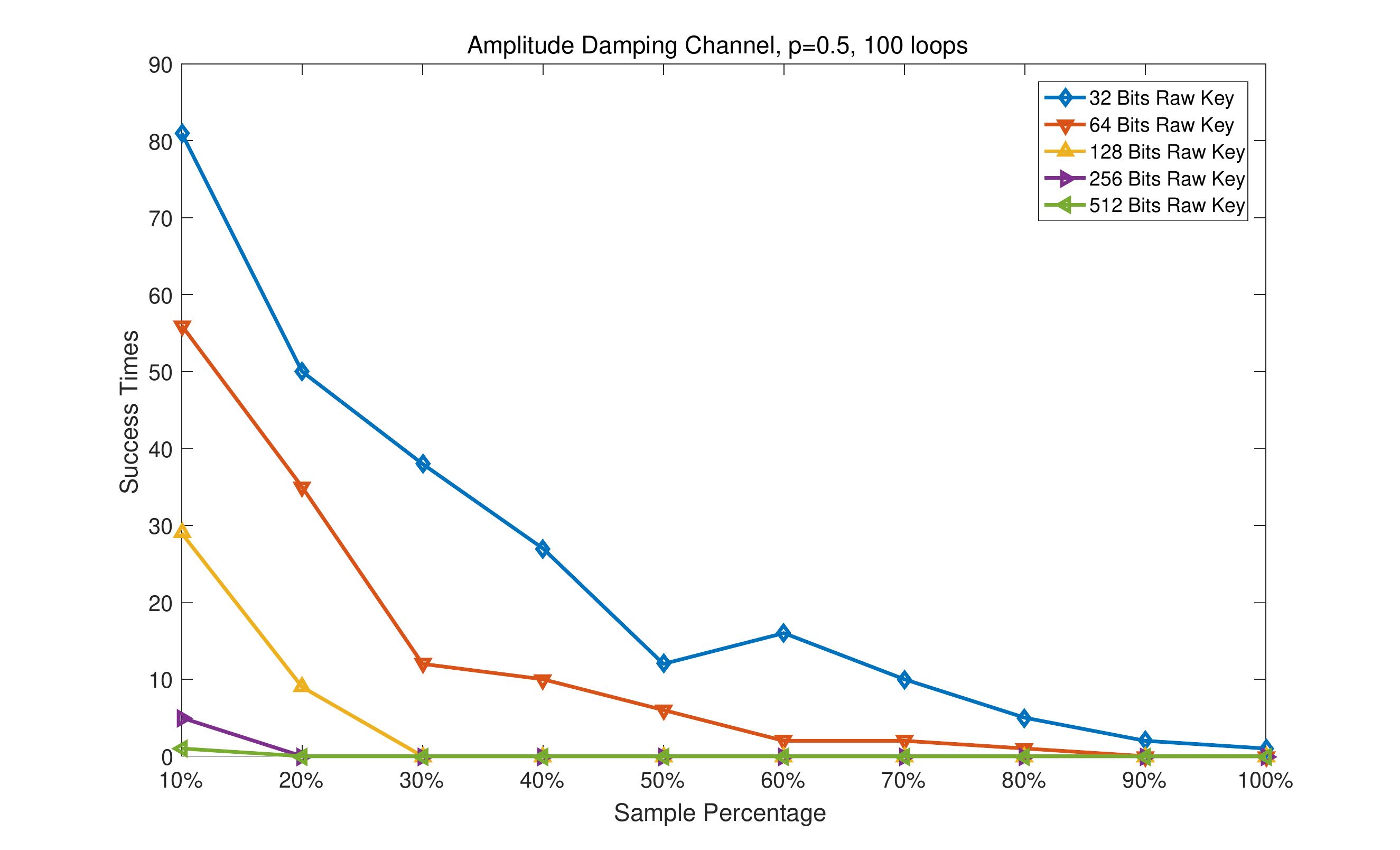}}
        \centerline{(f) Amplitude Damping Channel, $p=0.5$, $loops=100$}    
    \end{minipage}
    %\end{tabular}
    \caption{Statistics of success communication via BB84 with channels}
    \label{fig:BB84channel}
\end{figure}

Success times for different sampling percentages in different channels in 100 shots are provided by Fig.~\ref{fig:BB84channel}.

\subsubsection{Features and Analysis}

The example generates some `erroneous' bits during communication due to the quantum channels. These bits cause a connection failure. Meanwhile, not all error bits can be found in the sampling step because, in theory, almost half the bits are invalid in the measurement step. Additionally, the sampling step is also a probability verification step which means it does not use all the agreed bits to verify the communication procedure.

Subfigures (a),(b) and (c) in Fig.~\ref{fig:BB84channel} are bit flip channels with different probabilities. Overall, successful shots increase as $p$ increases and the raw key length shortens. This is because $p$ is a reflection of the percentage of information that remains in bit flip channel and an increase in $p$ means fewer errors in communication. A shorter raw key length ensures fewer bits are sampled.
Sub-figure (d),(e) and (f) illustrate the communication capacity of the BB84 protocol in the other three channels. Note that the $I$-identity channel has a 100\% success rate, which means it is a noiseless channel and can keep information intact during the transfer procedure.

\section{Grover's search algorithm}
Grover's search algorithm is a well-known quantum algorithm. It solves searching problems in databases consisting of $N$ elements, indexed by number $0,1,\ldots,N-1$ with an oracle provides the answer as a position. This algorithm can find solutions with a probability of $O(1)$ within $O(\sqrt{N})$ steps.

\subsection{A Simple Grover's search algorithm}
In this example, we assume there is only one answer to the question, i.e., the oracle will only reverse one position at a time. Further, the oracle is assumed to be working as a black box and can reverse the correct position of the answer. After querying the oracle $r=\frac{\pi}{4}\sqrt{N}$ times with the corresponding phase rotations, the quantum state includes the correct information to answer the question.
\subsubsection{Input and output}
{Input:}
\begin{itemize}
    \item The total number of space $N$. For convenience, we restrict $N=2^n$.
    \item The correct position of the search. That is used to construct oracle. 
\end{itemize}

{Output:}
\begin{itemize}
    \item The final position of the measurement result.
    \item Oracle time $r$.
\end{itemize}
\subsubsection{Results}
The simple Grover's search algorithm has only one result, and the final measurement result shows the correct answer to the searching problem.

\subsubsection{Features and analysis}
Suppose $\ket{\alpha}=\frac{1}{\sqrt{N-1}}\sum_{x}^{''}\ket{x}$ is not the solution but rather $\ket{\beta}=\sum_{x}^{'}\ket{x}$ is the solution where $\sum_{x}^{'}$ indicates the sum of all the solutions. The initial state $\ket{\psi}$ may be expressed as
\[
\ket{\psi}=\sqrt{\frac{N-1}{N}}\ket{\alpha}+\sqrt{\frac{1}{N}}\ket{\beta}\, .
\]
Every rotation makes the $\theta$ to the solution where
\[
\sin \theta=\frac{2\sqrt{N-1}}{N}\, .
\]
When $N$ is larger, the gap between the measurement result and the real position number is less than $\theta=\arcsin \frac{2\sqrt{N-1}}{N}\approx \frac{2}{\sqrt{N}}$. Therefore, it is almost impossible to have a wrong answer within $r$ times.

\subsection{Multi-object Grover's search algorithm}
A more general Grover's search algorithm is considered: a multi-object Grover's search algorithm. This case supposes that there may be more than one correct answer (position) for the oracle to find. We use a strategy that adds a blind box to reverse the proper position of the answer. This experiment reveals that Grover's algorithm leads to an avalanche of error in a multi-object setting, indicating that algorithm needs to be modified in some way.

A new blind box (a unitary gate) is added, which reverses the proper position of the answer. In short, the oracle is a matrix where all the diagonal elements are $1$, but all the answer positions are $-1$.  Thus, the blind box is a diagonal matrix where all elements are $1$, and all the answer position that have been found are $-1$. When these two boxes are combined, we create a new oracle with the answers to all the questions except ones were found in previous rounds.
\subsubsection{Input and output}

The input is
\begin{itemize}
    \item The total number of spaces $N$. For convenience, we restrict $N=2^n$.
    \item All correct positions of the search. 
\end{itemize}

The output is
\begin{itemize}
    \item The final position of the measurement result.
    \item Oracle time $r$.
\end{itemize}
\subsubsection{Results}
The measurement shows different probabilities of the final result. The theory holds that if we have multiple-answers, the state after $r$ times oracles and phase gates become the state near both of them.  For example, if the answers are $\ket{2}\, ,\ket{14} \in \cH_{64}$, the state before the measurement is expected to be almost  $\frac{1}{\sqrt{2}}(\ket{2}+\ket{14})$. We should get $\ket{2}$ or $\ket{14}$ the first time and the other one the next time. However, we get results other than $\ket{2}$ and $\ket{14}$ with high probability, which indicates that the multi-object search algorithm is not very good.
 
\subsubsection{Features and analysis}
It worth noting that due to multi-objects, the real state after using Grover's search algorithm becomes  $a(\ket{2}+\ket{14})+b(\ket{1}+\ket{3}+\ket{4}+\ket{5}+....)$ where $a,b \in \mathcal{C}$ and $|a|^2+|b|^2=1$. However, $b$ cannot be ignored even it is very small. An interesting issue occurs when the wrong position index is found. If the wrong index is measured, the algorithm creates an incorrect blind box and reverses the wrong position of the oracle, i.e., it adds a new answer to the questions. In next round, the proportion of correct answers is further reduced. In the last example, we would have gotten a wrong answer by measurement, say $\ket{5}$. After new procedure, the state would become: $a(\ket{2}+\ket{14}+\ket{5})+b(\ket{1}+\ket{3}+\ket{4}+\ket{5}+....)$. It becomes harder and harder to find the correct answer with this state. 
%In conclusion, one wrong answer in a round causes an avalanche of errors with Grover's search algorithm.
\end{document}